\documentstyle[aas2pp4]{article}

\lefthead{Nakano and Makino}
\righthead{Origin of Density Cusps in Elliptical Galaxies}

\begin{document}

\title{On the Origin of Density Cusps in Elliptical Galaxies}

\author{Taro Nakano and Junichiro Makino}
\affil{Department of General Systems Studies, College of Arts and
Sciences, University of Tokyo, \\
3-8-1 Komaba, Meguro-ku, Tokyo 153, Japan}

\begin{abstract}
We investigated the dynamical reaction of the central region of
galaxies to a falling massive black hole by $N$-body simulations. As
the initial galaxy model, we used an isothermal King model and placed
a massive black hole at around the half-mass radius of the galaxy. We
found that the central core of the galaxy is destroyed by the heating
due to the black hole and that a very weak density cusp ($\rho \propto
r^{-\alpha}$, with $\alpha \sim 0.5$) is formed around the black hole. 
This result is consistent with recent observations of large elliptical 
galaxies with {\it Hubble Space Telescope}. The velocity of the stars
becomes tangentially anisotropic in the inner region, while in the
outer region the stars have radially anisotropic velocity
dispersion. The radius of the weak cusp region is larger for larger
black hole mass. Our result naturally explains the formation of the
weak cusp found in the previous simulations of galaxy merging, and
implies that the weak cusp observed in large elliptical galaxies may
be formed by the heating process by sinking black holes during merging
events.
\end{abstract}


\section{Introduction}

Recent {\it HST\/} observations (\cite{lau95}; \cite{byu96};
\cite{geb96}; \cite{fab96}; \cite{kor96}) have drastically changed our
understanding of the structure of elliptical galaxies. First, no
elliptical galaxies have cores with constant surface brightness. All
of them turned out to have cusps with power-law surface brightness
profiles. Second, elliptical galaxies are divided into two groups in
terms of the slope of the cusp: ``weak-cusp'' galaxies and
``strong-cusp'' galaxies. The weak-cusp galaxies have the surface
brightness profiles expressed as $I(R) \propto R^{-\gamma}$ with $0
\lesssim \gamma \lesssim 0.3$ and deprojected luminosity density
profiles expressed as $\rho (r) \propto r^{-\alpha}$ with $0.3
\lesssim \alpha \lesssim 1.1$ (\cite{kor96}). On the other hand, the
cusps of the strong-cusp galaxies are expressed by the same formula,
but with $\gamma \simeq 1$ and $\alpha \simeq 2$. In addition, it was
found that bright galaxies have weak cusps, while faint galaxies have
strong cusps.

Previously, bright elliptical galaxies were believed to have cores
with constant surface brightness. High-resolution surface photometry
showed that these cores are non-isothermal (\cite{lau85};
\cite{kor85}). {\it HST} observations revealed these cores are actually
weak cusps. 

The origin of such weak cusps is not well understood. One possible way
to form a density cusp is to introduce the central black
hole. Young (1980) studied the response of the structure of a stellar
system to the adiabatic growth of the central black hole. He showed
that the power-law density cusp with $\rho \propto r^{-3/2}$ is formed
around the black hole. The same result was obtained by Quinlan,
Hernquist, \& Sigurdsson (1995).

Another way to form cusps using a central black hole was proposed by
Bahcall \& Wolf (1976). They investigated the evolution of the
structure of a stellar system around a black hole driven by two-body
relaxation. They found a steady state solution with a steep
cusp. Shapiro \& Marchant (1978) and Marchant \& Shapiro (1979, 1980)
performed Monte Carlo simulation and found that this steady state cusp
is actually formed.

Both models predict cusps steeper than the observed ones. Therefore
the origin of these weak cusps cannot be explained by means of these
mechanisms.

Navarro, Frenk, \& White (1996) found that the dark matter halos formed 
in standard CDM cosmogony is well approximated by a ``universal
profile'' which approaches $\rho \propto r^{-1}$ near the
center. They argued that this result might explain the origin of weak
cusps. Recently, Fukushige \& Makino (1997) performed the same
calculation as Navarro et al. (1996), but with much higher
resolution in both mass and space. They obtained a nearly isothermal
cusp. Thus the cusps formed by CDM clustering scenario are also too
steep.

So far, the only scenario that successfully reproduced the weak cusps
is the merging of two galaxies with central black holes (\cite{mak96},
hereafter ME). They investigated the mergings of two galaxies with
central black holes by {\it N}-body simulations and found that the
merger remnants have very shallow ($\rho \propto r^{-1}$ or shallower)
density cusps in the central regions. This result is in good agreement
with the {\it HST\/} observations.

The physical process by which the weak cusps are formed in a merger
remnant was not clearly explained in ME. It is obvious that the black
holes played some important roles in the cusp formation, since no
significant cusps were formed by the merger remnant of two galaxies
without black holes. The merging of two galaxies with black holes
involves several processes, such as the sinking of two black holes to the
center of the remnant due to dynamical friction from field stars, the
formation and hardening of a black hole binary and its back reaction to
field stars. We need simpler models with which we can study
the effect of each elementary process separately.

In the present work, we study the sinking of a black hole due to
dynamical friction and its back reaction to the distribution of field
stars. We performed $N$-body simulation of the system of spherical
galaxy model and a black hole placed at around the half mass radius of 
the galaxy. We found the shallow density cusp with $\rho \sim
r^{-0.5}$ is formed in the center of galaxies. Thus we can conclude
that the sinking of black hole is the dominant mechanism of the
formation of weak cusps. 

In the next section, we describe the numerical methods and initial
models we used. We present our results in \S \ref{sec:result} and
a discussion and summary in \S \ref{sec:summary}.

\section{Initial Conditions and Numerical Method}
\subsection{Initial Conditions}
\begin{figure}[htbp]
\plotone{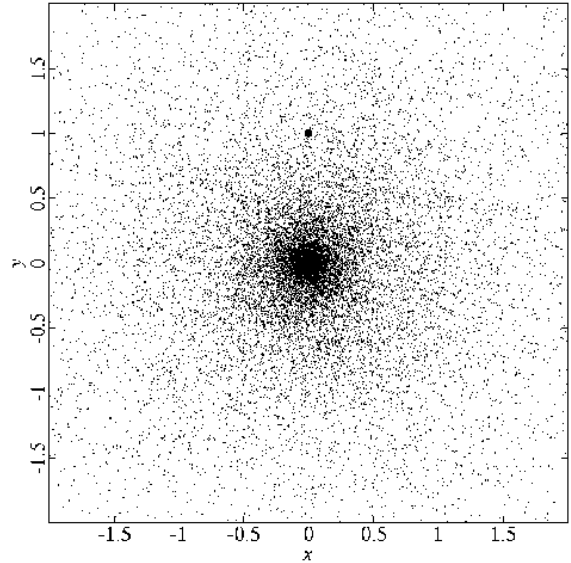}
\caption{The snapshot of the whole system at $t=0$ for run A. The %
filled circle on $(x, y)=(0, 1)$ shows the initial position of a %
massive BH. \label{fig:1}}
\end{figure}

The initial conditions we used are summarized in Table
\ref{tab:1}. Column 2 shows the initial galaxy models. We used
King models with the central potential parameter $W_{\rm c}=9$, 6 and
3 as initial galaxy models. The number of particles in the initial
galaxies (column 3) was 32767 or 131071, and all particles have equal
mass. We adopted the standard unit (\cite{heg86}) as the
system of units, in which $G=M_{\rm gal}=1$ and $E_{\rm gal}=-1/4$,
where $G$ is the gravitational constant, $M_{\rm gal}$ is the total
mass of the initial galaxy and $E_{\rm gal}$ is the total energy of
the galaxy. In these units, the half-mass radii of the initial
galaxies are 0.98, 0.80 and 0.84 for the King models with $W_{\rm
c}=9$, 6 and 3, respectively. The half-mass crossing time is
$2\sqrt{2}$ for all models. Column 4 shows the black hole masses. We
used different black hole masses (2, 4, 8 \% of the mass of the whole
system). Column 5 shows the initial position and velocity of the black
hole. We tried three different set of initial conditions of the black
hole. In ``off-center'' runs (runs A, F, G, H, I and J), we initially
placed a black hole in the position of $r=1$ from the center of mass
of the galaxy (Figure \ref{fig:1}). In the ``spiral-in'' runs (run B,
C and D), we placed a black hole in the same position as in the
off-center runs with some initial tangential velocity $v_{\rm t}$. We
varied $v_{\rm t}$ as $v_{\rm K}$, $0.5 v_{\rm K}$ and $0.25 v_{\rm
K}$, where $v_{\rm K}=\left. \sqrt{G M(r)/r} \; \right| _{r=1}$ (circular
velocity at the initial position). In the third set, which is the
``on-center'' run (run E), however, we put the black hole exactly at
the center of the galaxy, to compare the result with those of other
runs. For both off-center and on-center runs, the initial velocity of
the black hole was zero.

\placetable{tab:1}

\begin{deluxetable}{ccccc}
\tablewidth{0pt}
\tablecaption{Initial Conditions \label{tab:1}}
\tablehead{
\colhead{Run} & \colhead{Galaxy Model} & \colhead{$N$} &
\colhead{$M_{\rm BH}$} & \colhead{BH Initial Condition} \\
\colhead{(1)} & \colhead{(2)} & \colhead{(3)} &
\colhead{(4)} & \colhead{(5)}
}

\startdata
A & King ($W_{\rm c}=9$) & 32767 & 1/24 & $r=1$, $v_{\rm t}=0$ \nl
B & King ($W_{\rm c}=9$) & 32767 & 1/24 & $r=1$, $v_{\rm t}=v_{\rm K}$ \nl
C & King ($W_{\rm c}=9$) & 32767 & 1/24 & $r=1$, $v_{\rm t}=0.5 v_{\rm K}$ \nl
D & King ($W_{\rm c}=9$) & 32767 & 1/24 & $r=1$, $v_{\rm t}=0.25 v_{\rm K}$ \nl
E & King ($W_{\rm c}=9$) & 32767 & 1/24 & $r=0$ \nl
F & King ($W_{\rm c}=9$) & 32767 & 1/49 & $r=1$, $v_{\rm t}=0$ \nl
G & King ($W_{\rm c}=9$) & 32767 & 2/23 & $r=1$, $v_{\rm t}=0$ \nl
H & King ($W_{\rm c}=6$) & 32767 & 1/24 & $r=1$, $v_{\rm t}=0$ \nl
I & King ($W_{\rm c}=3$) & 32767 & 1/24 & $r=1$, $v_{\rm t}=0$ \nl
J & King ($W_{\rm c}=9$) & 131071 & 1/24 & $r=1$, $v_{\rm t}=0$ \nl
\enddata
\end{deluxetable}

\placefigure{fig:1}

\subsection{Numerical Method}
We used the {\tt NBODY1} calculation code (\cite{aar85}) modified for
GRAPE-4, a special-purpose computer for astronomical $N$-body
problem (\cite{tai96}). In this code, we used the Hermite
scheme (\cite{mak92}) to carry out time integration of
orbits. The equation of motion for star $i$ is given by
\begin{equation}
\frac{d^{2} \mbox{\boldmath $r$}_{i}}{d t^{2}}=-\sum_{j \neq i}\frac{G
m_{j} (\mbox{\boldmath $r$}_{i} - \mbox{\boldmath
$r$}_{j})}{(|\mbox{\boldmath $r$}_{i} - \mbox{\boldmath $r$}_{j}|^{2} +
\epsilon^{2})^{3/2}}
\end{equation}
where \mbox{\boldmath $r$} and $m$ are the position and mass of the
star, respectively, and $\epsilon$ is the softening parameter which is
added to avoid divergence of the gravitational force between close
stars. The softening length is 1/1024 for the force between field star
particles and is $10^{-8}$ for the force between a black hole particle
and a field star. These softening parameters are small enough
to achieve necessary spatial resolution, since we are interested in the
central structure of the scale of $\sim 10^{-2}$ or larger. The
maximum relative energy error at the end of the run was $1.079 \times
10^{-4}$ (for run A) and typical error was $10^{-5}$ along the
calculations. The black hole is represented by a particle that has
much larger mass than that of a field particle. The mass accretion to
the black hole through tidal disruption of field stars or relativistic
effects were not taken into account.

\section{Results} \label{sec:result}

\subsection{Motion of the Black Hole} \label{subsec:motionbh}

\begin{figure}[htbp]
\plotone{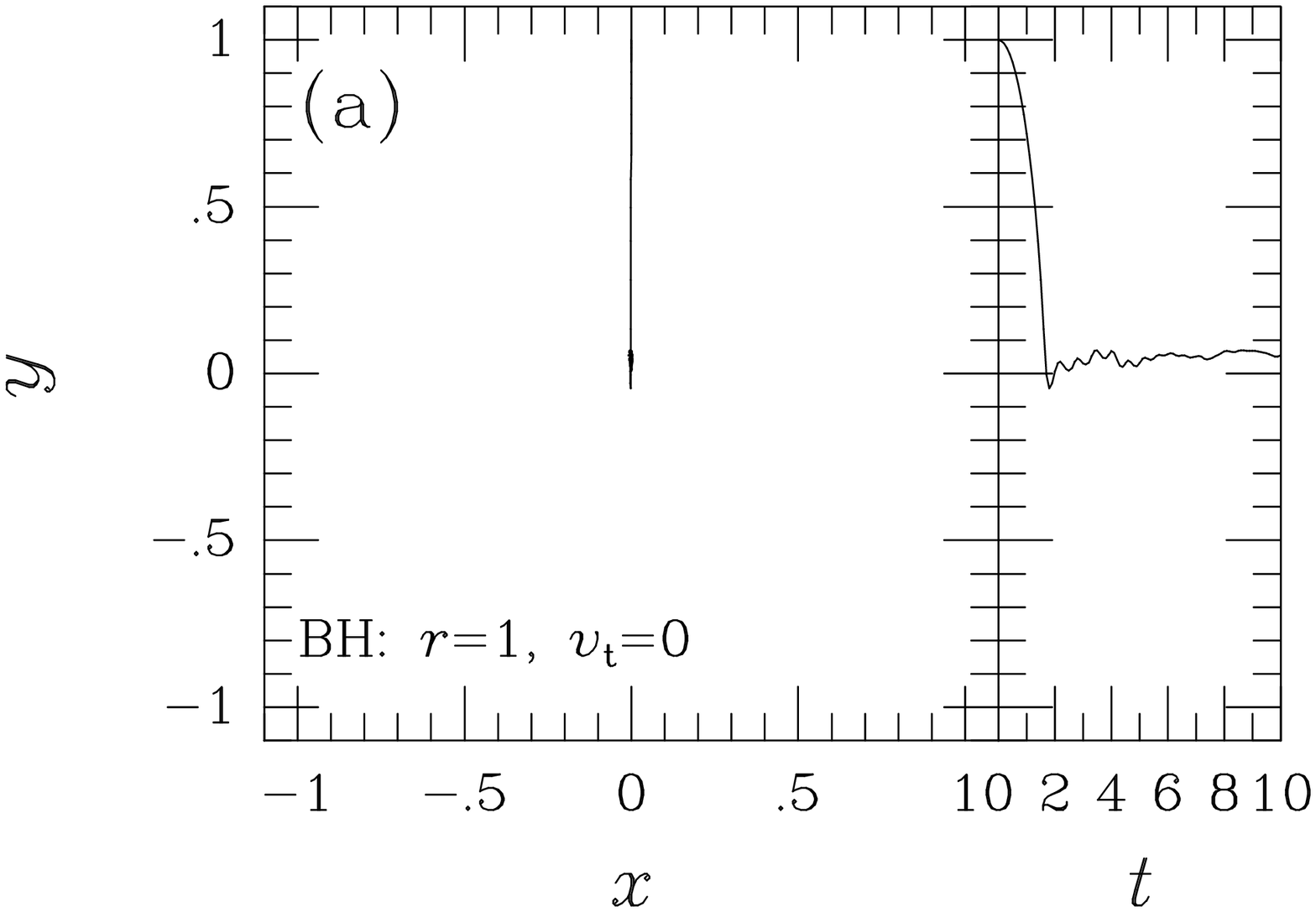}\\
\plotone{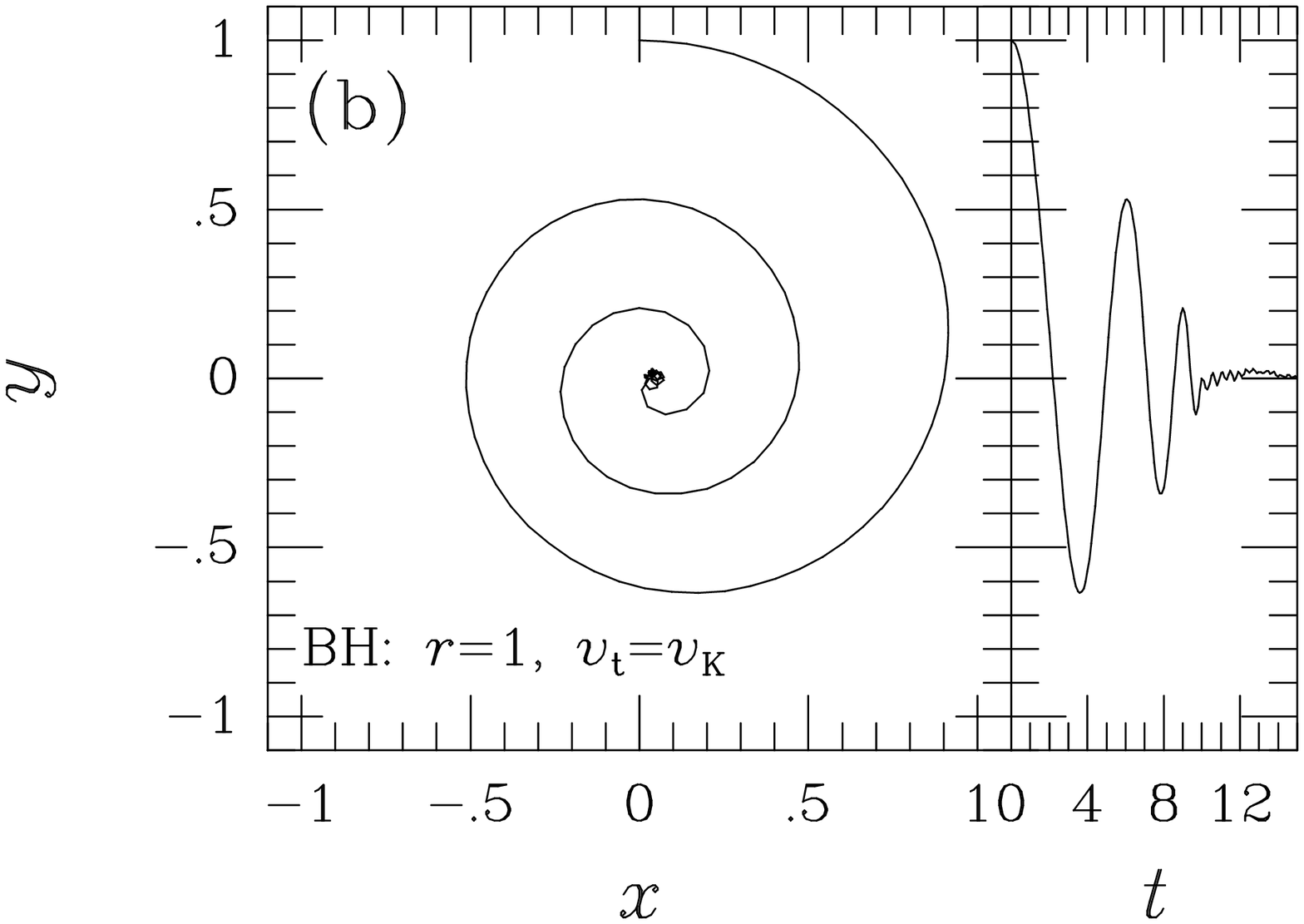}\\
\plotone{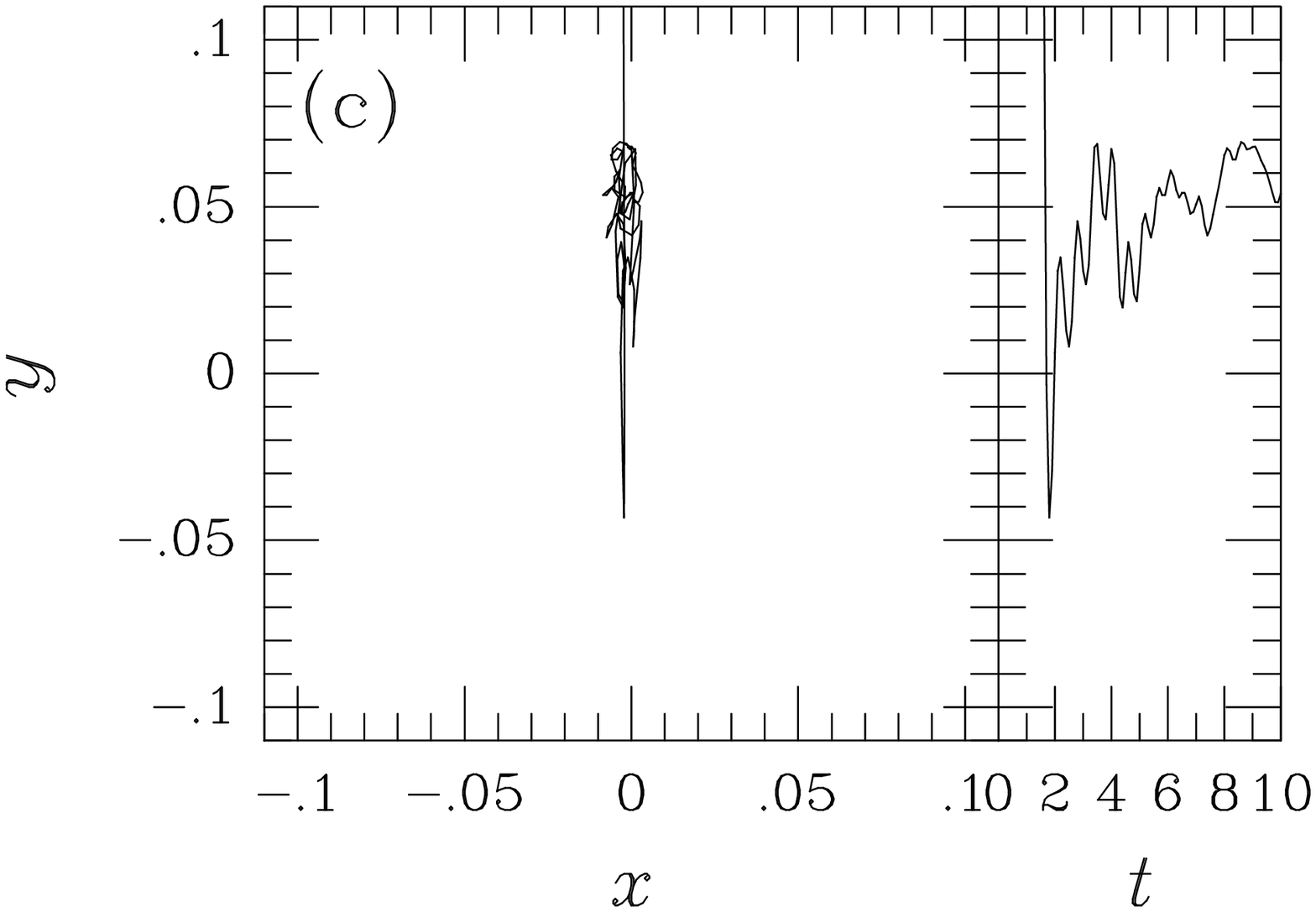}
\caption{(a) Left: Trajectory of a BH on $x-y$ plane for %
$t=0-10$ for off-center BH (run A); Right: Time evolution of the %
$y$-component of the BH's position. (b) Same as (a) but for spiral-in %
BH (run B). (c) 10 times magnified view of the central region of
(a). \label{fig:2}}
\end{figure}

Figure \ref{fig:2} shows the trajectory of the black hole in the
center-of-mass reference frame for the off-center run (run A) and the
spiral-in run (run B). The time ranges from $t=0$ to $t=10$ for run A
and from $t=0$ to $t=15$ for run B. In this coordinate system, the
massive black hole is initially put on $(x, y, z)=(0, 1, 0)$. In the
off-center run (Figure \ref{fig:2}a and c), the black hole
oscillates around the center of the system and its amplitude is damped
quickly due to the dynamical friction of the field stars. In the
spiral-in run (Figure \ref{fig:2}b), the black hole sinks into the
center along the spiral orbit. Even after the oscillation or spiral
motion is damped ($t \gtrsim 5$ for run A and $t \gtrsim 10$ for run
B), the black hole moves randomly around the bottom of the potential
well, because the random velocity of the black hole reaches
equipartition with that of field stars. The typical length scale of
this random motion is, from Figure \ref{fig:2}c, around 0.01. Thus, we
can analyze the system, assuming that the black hole is at the center,
up to the radius of 0.01. The all following results in this paper are
described on this assumption. The structure smaller than this radius
is probably meaningless.  

\placefigure{fig:2}

\subsection{Density Profiles}
\begin{figure}[htbp]
\plotone{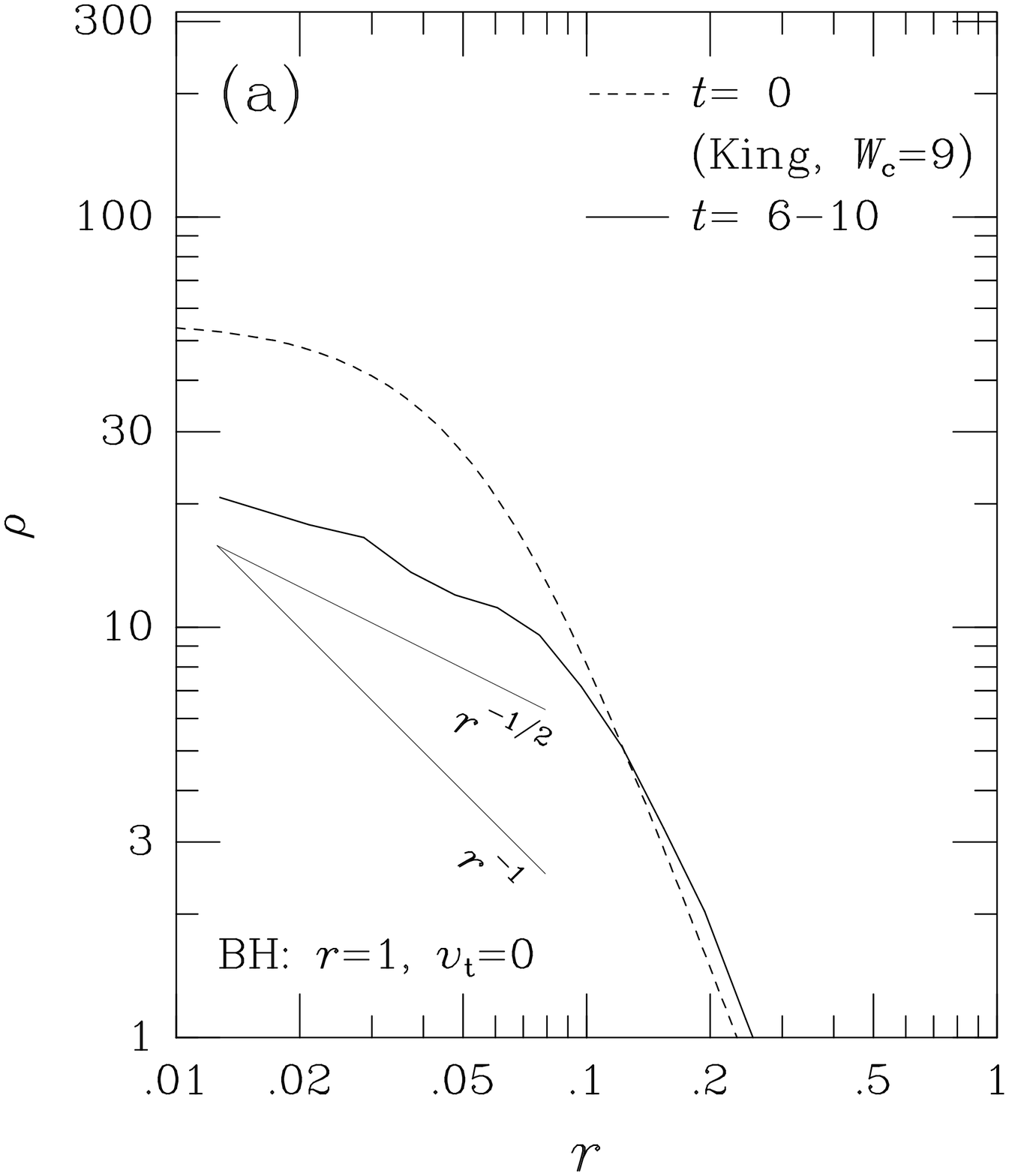}\\
\plotone{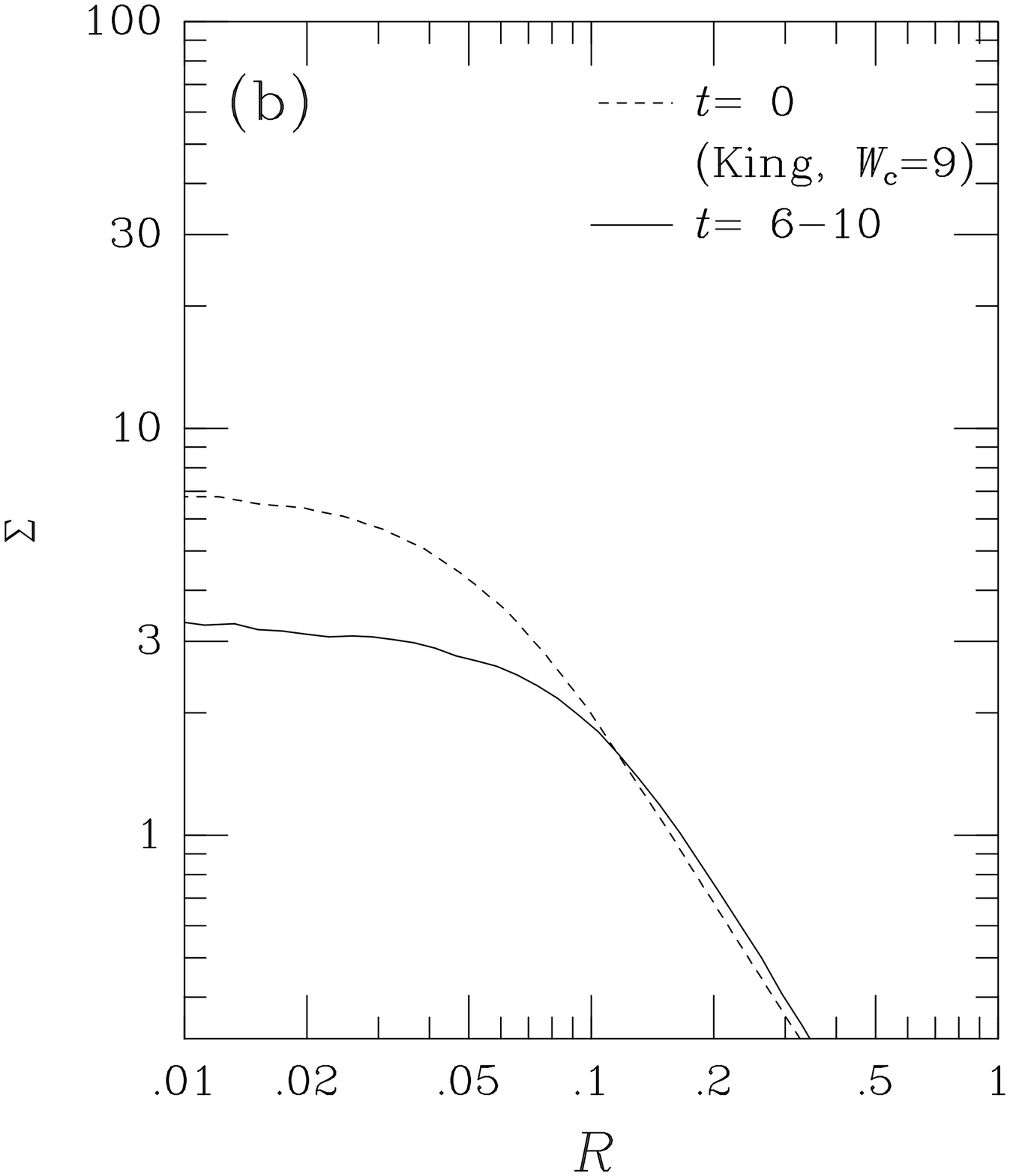}
\caption{Results for off-center BH (run A): (a) density profile; %
(b) surface density profile. \label{fig:3}}
\end{figure}

\begin{figure}[htbp]
\plotone{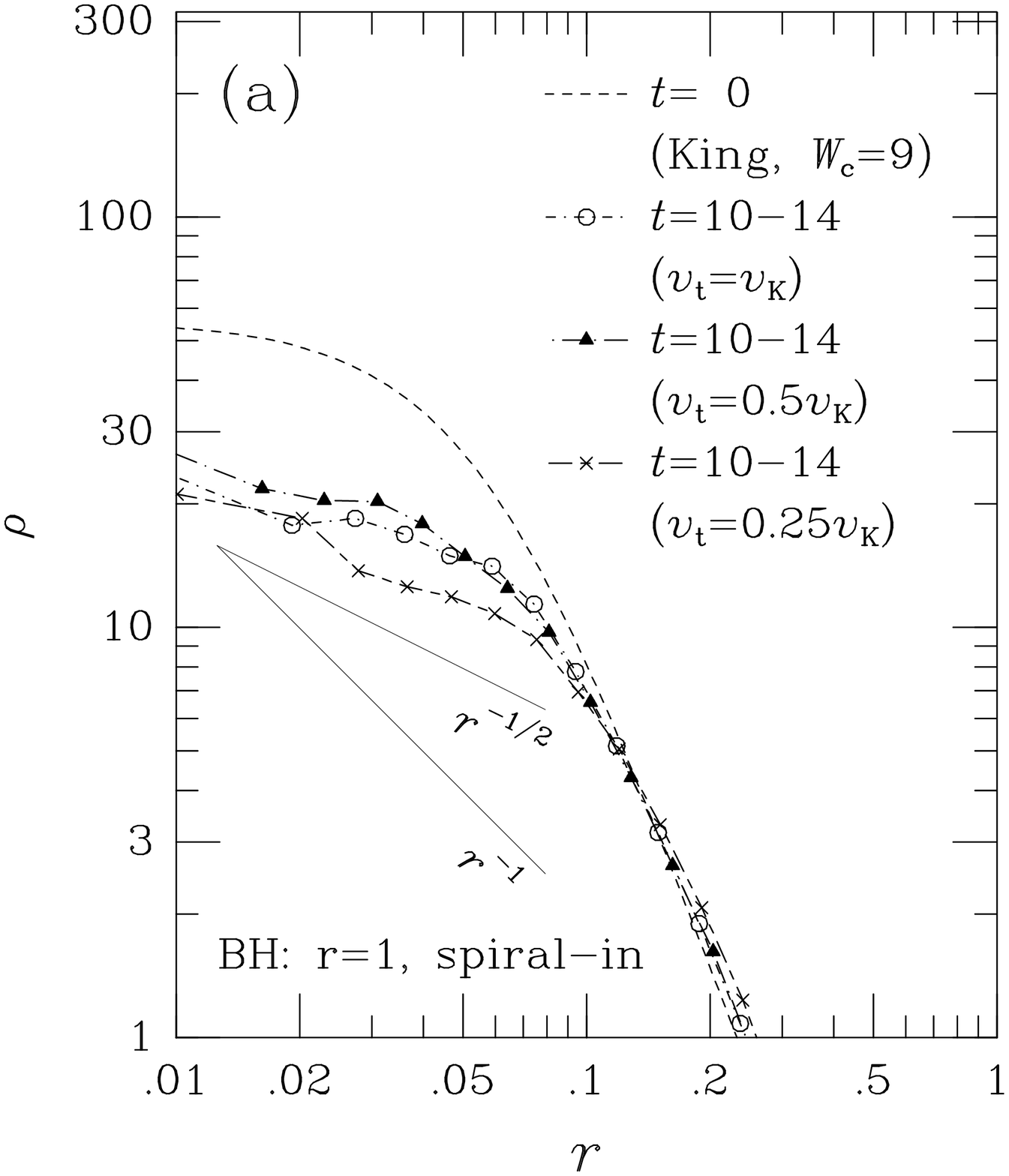}\\
\plotone{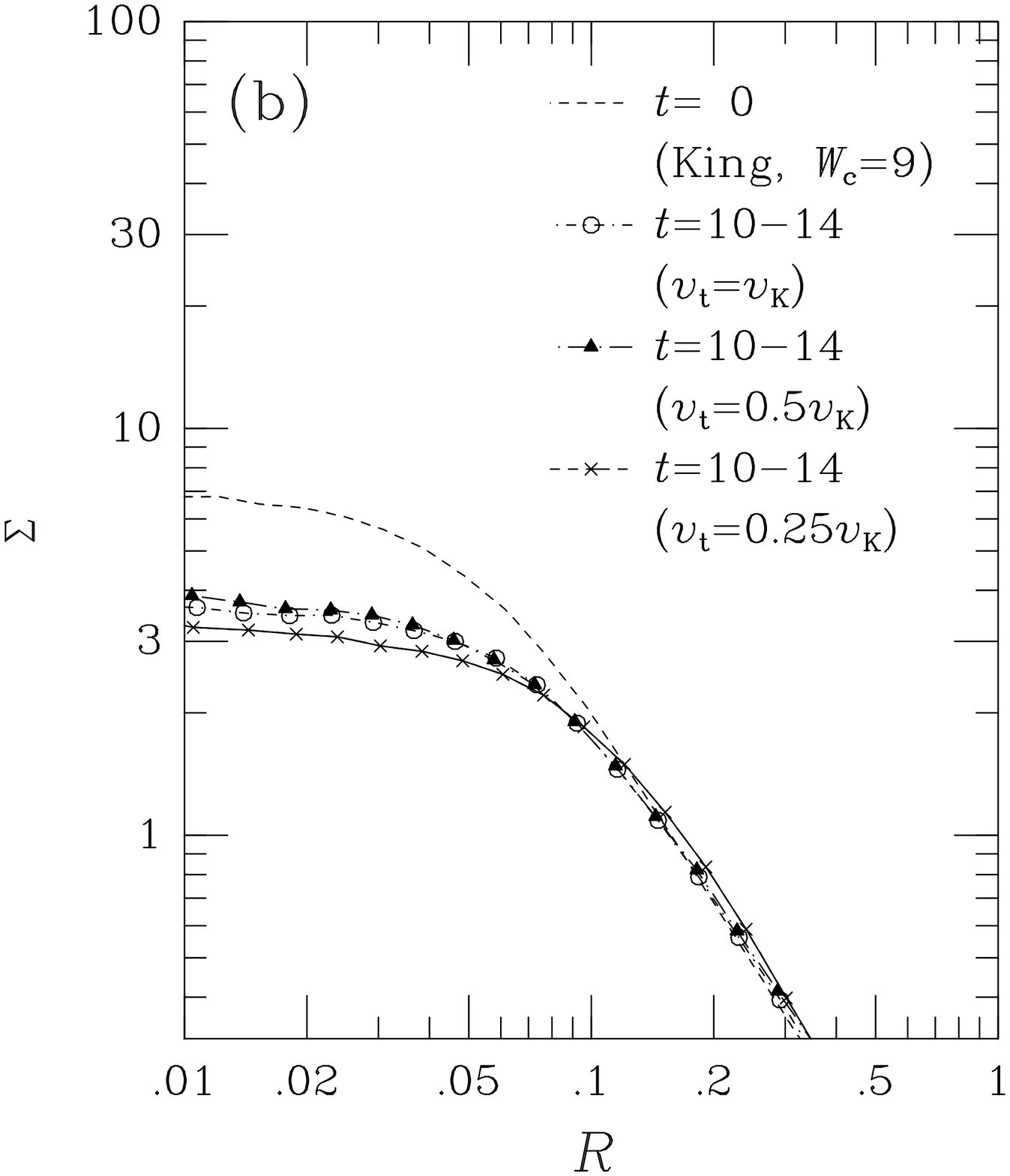}
\caption{(a),(b) Same as Figure \protect\ref{fig:3} but for
spiral-in BH (run B, C and D). \label{fig:4}}
\end{figure}

\begin{figure}[htbp]
\plotone{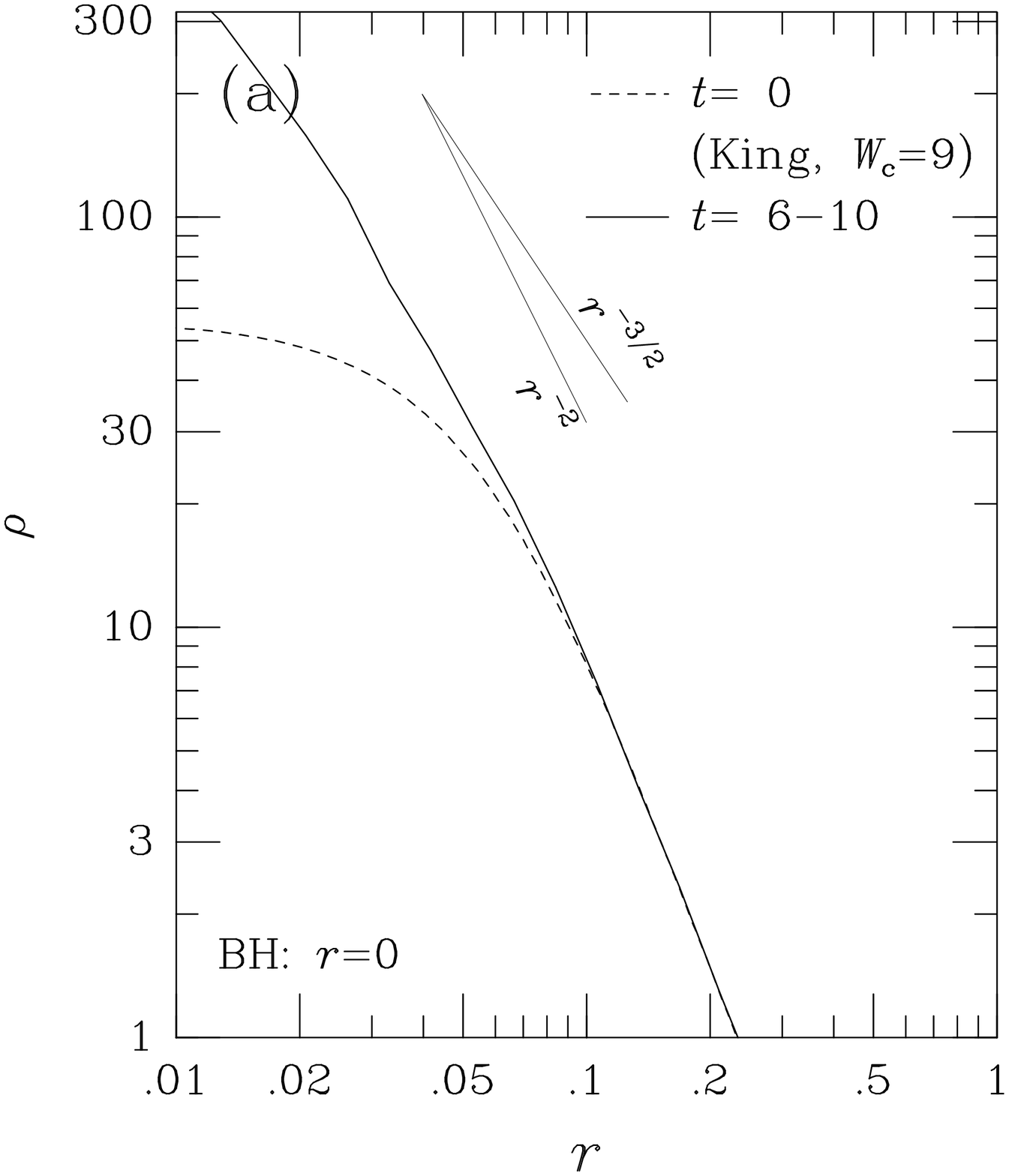}\\
\plotone{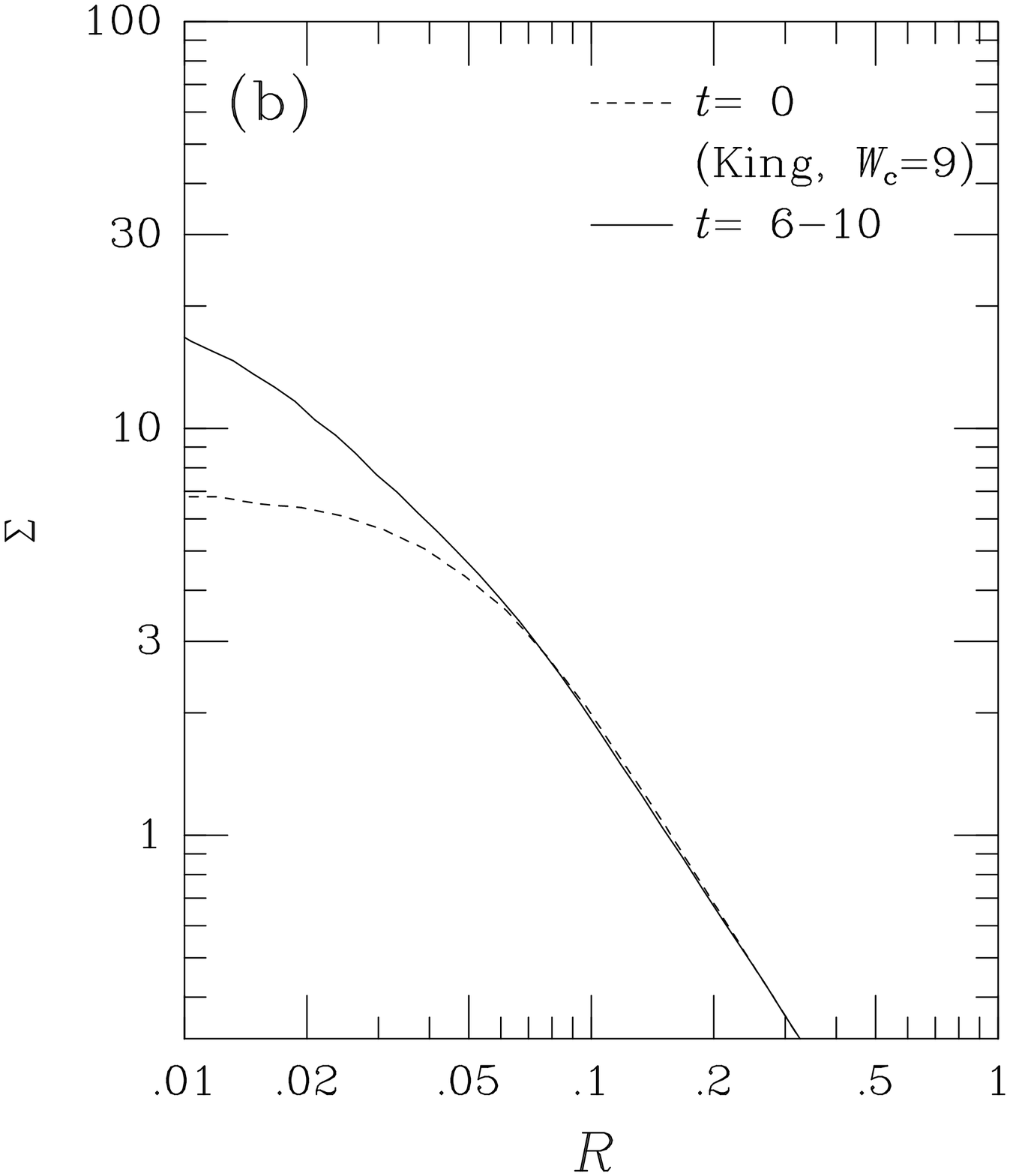}
\caption{(a),(b) Same as Figure \protect\ref{fig:3} but for
on-center BH (run E). \label{fig:5}}
\end{figure}

Figures \ref{fig:3}, \ref{fig:4} and \ref{fig:5} show the spatial and
surface density profiles for the off-center run (run A), the spiral-in
runs (runs B,C and D), and the on center run (run E). As mentioned in
\S\ref{subsec:motionbh}, we take the position of the black hole as the
center of galaxy (and thus as the origin of the coordinate system) in
the results of run A for $t \gtrsim 5$ and runs B, C and D for $t
\gtrsim 10$. We calculated the surface density $\Sigma (R)$ by summing
up the contribution of all stars outside radius $R$:
\begin{eqnarray}
\Sigma (R) &=& 2\int_{R}^{\infty}\frac{r\rho}{\sqrt{r^{2}-R^{2}}}dr
\nonumber \\
           &=& \frac{1}{2\pi}\sum_{i(r_{i}>R)}
               \frac{m_{i}}{r_{i}\sqrt{r_{i}^{2}-R^{2}}}, 
\end{eqnarray}
where $r_{i}$ and $m_{i}$ are the distance from the center of the galaxy 
and the mass of star $i$, respectively. 

In Figure \ref{fig:3}, we can see that the central region of the initial
King profile was destroyed by the black hole and a very weak cusp of
$\rho \propto r^{-0.5}$ is formed. This result is similar to that of
ME and in good agreement with the {\it HST\/} observations
(\cite{lau95}; \cite{byu96}; \cite{geb96}; \cite{fab96}; \cite{kor96}). For
the spiral-in runs (Figure \ref{fig:4}), we obtained the density
profiles similar to the off-center run. The central region has $\rho
\propto r^{-0.5}$ cusp for all cases. On the other hand, when the
black hole is placed initially at the center, the density structure
becomes quite different (Figure \ref{fig:5}a and b). The cusp in
Figure \ref{fig:5}a is approximately $\rho \propto r^{-1.5}$.

\placefigure{fig:3}

\placefigure{fig:4}

\placefigure{fig:5}

\subsection{Velocity Structure}
\begin{figure}[htbp]
\plotone{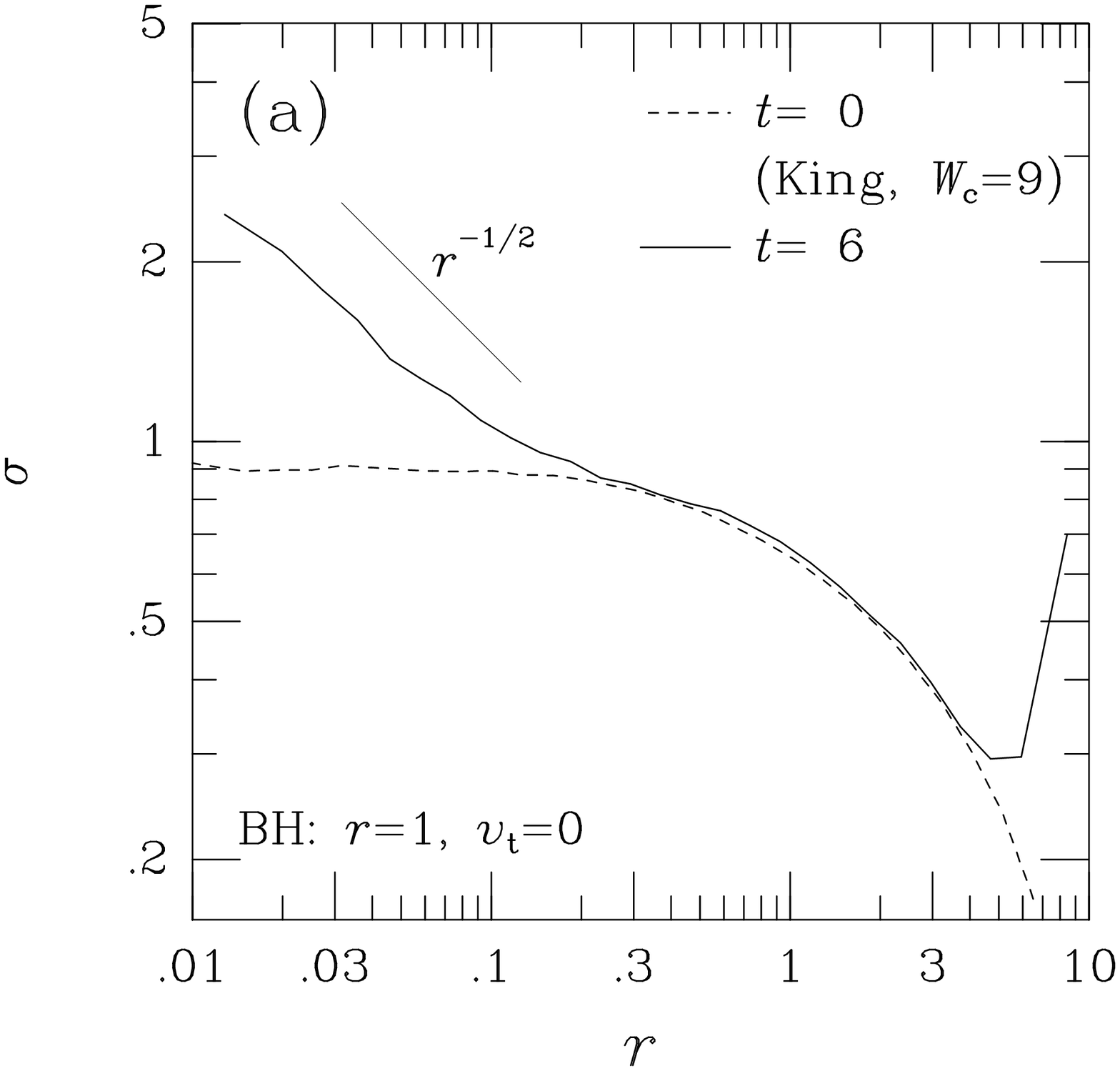}\\
\plotone{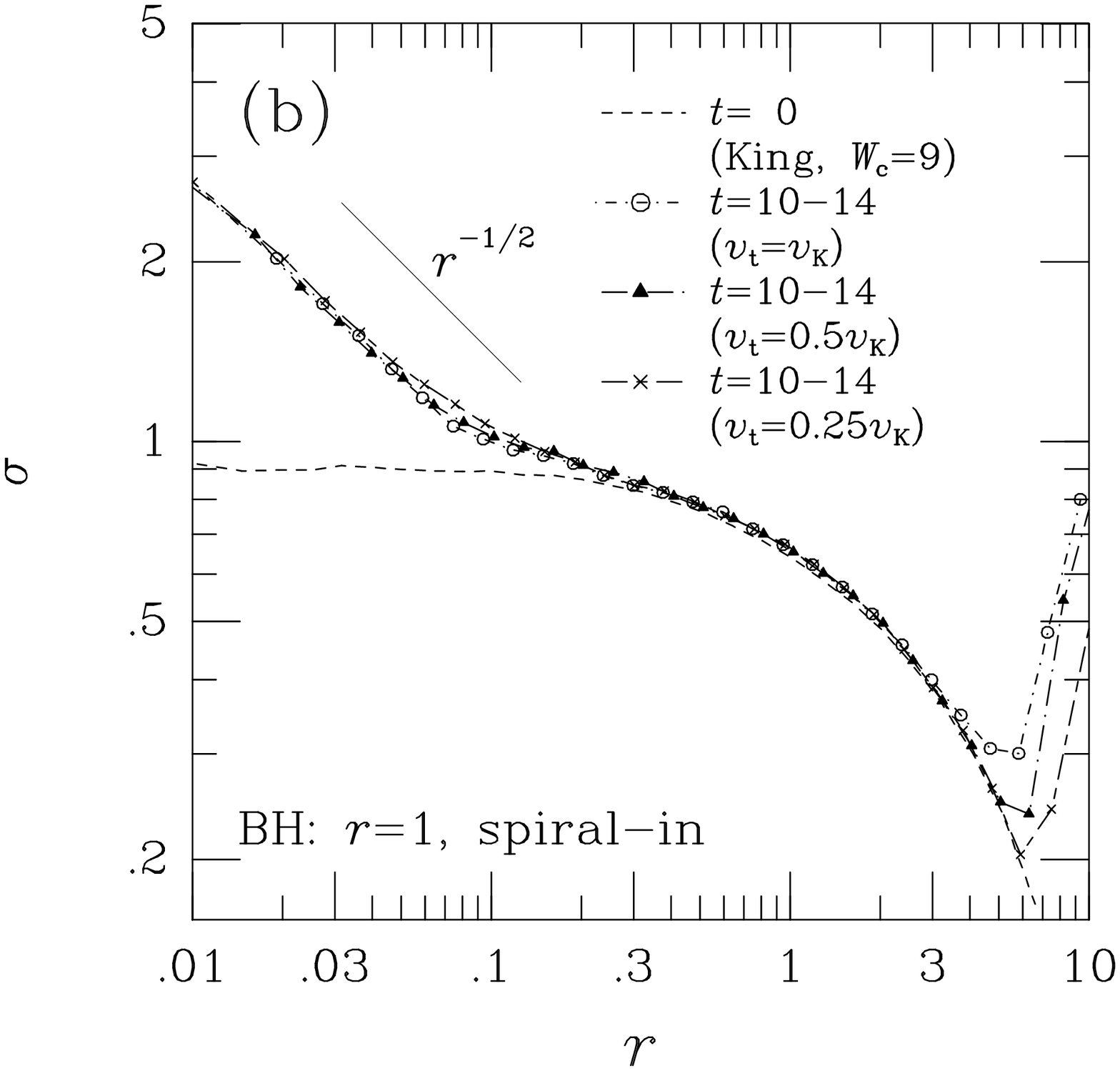}\\
\plotone{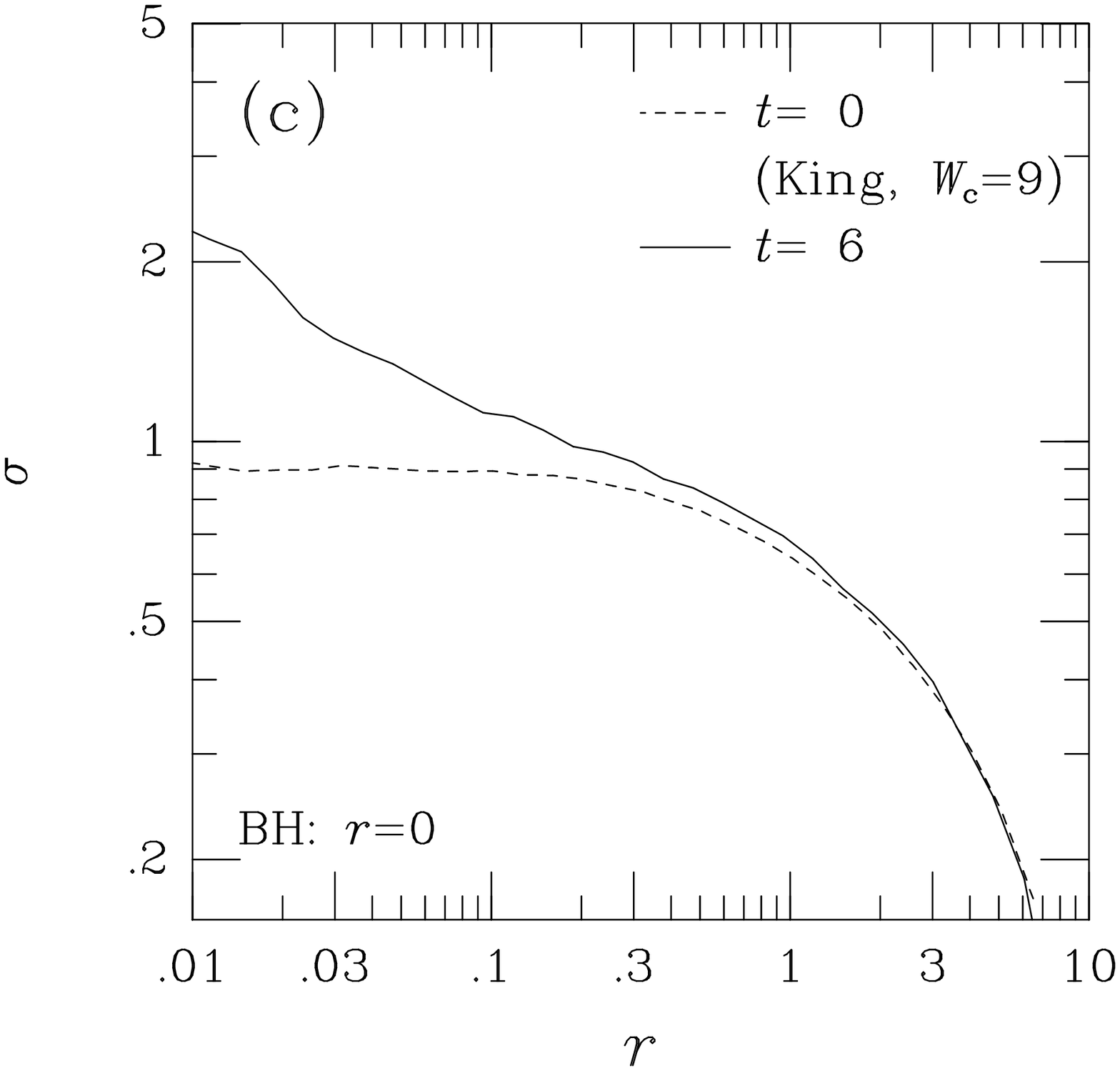}
\caption{Velocity dispersion profiles. (a) for off-center BH (run A); %
(b) for spiral-in BH (runs B, C and D); (c) for on-center BH (run %
E). \label{fig:6}}
\end{figure}

\begin{figure}[htbp]
\plotone{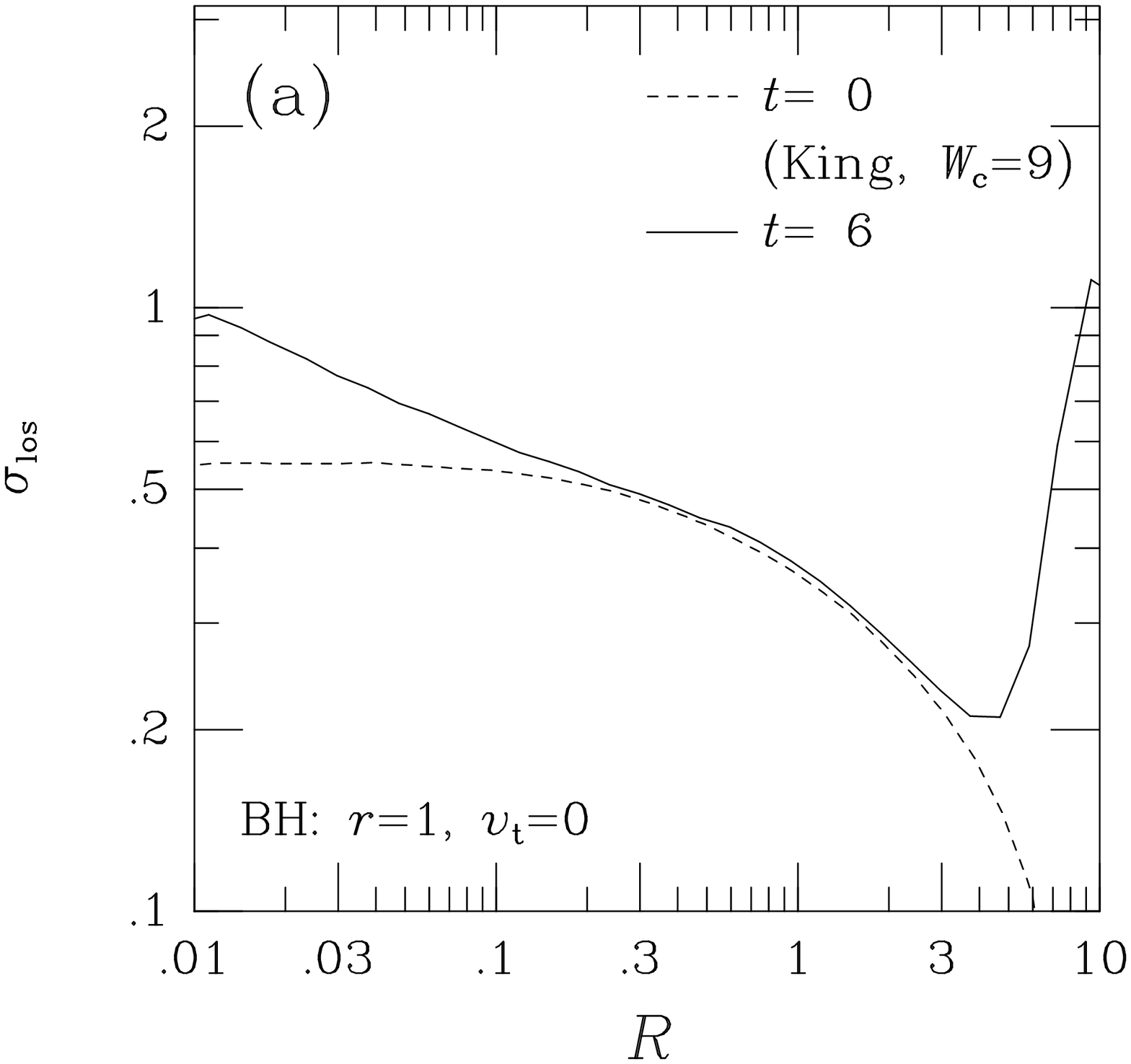}\\
\plotone{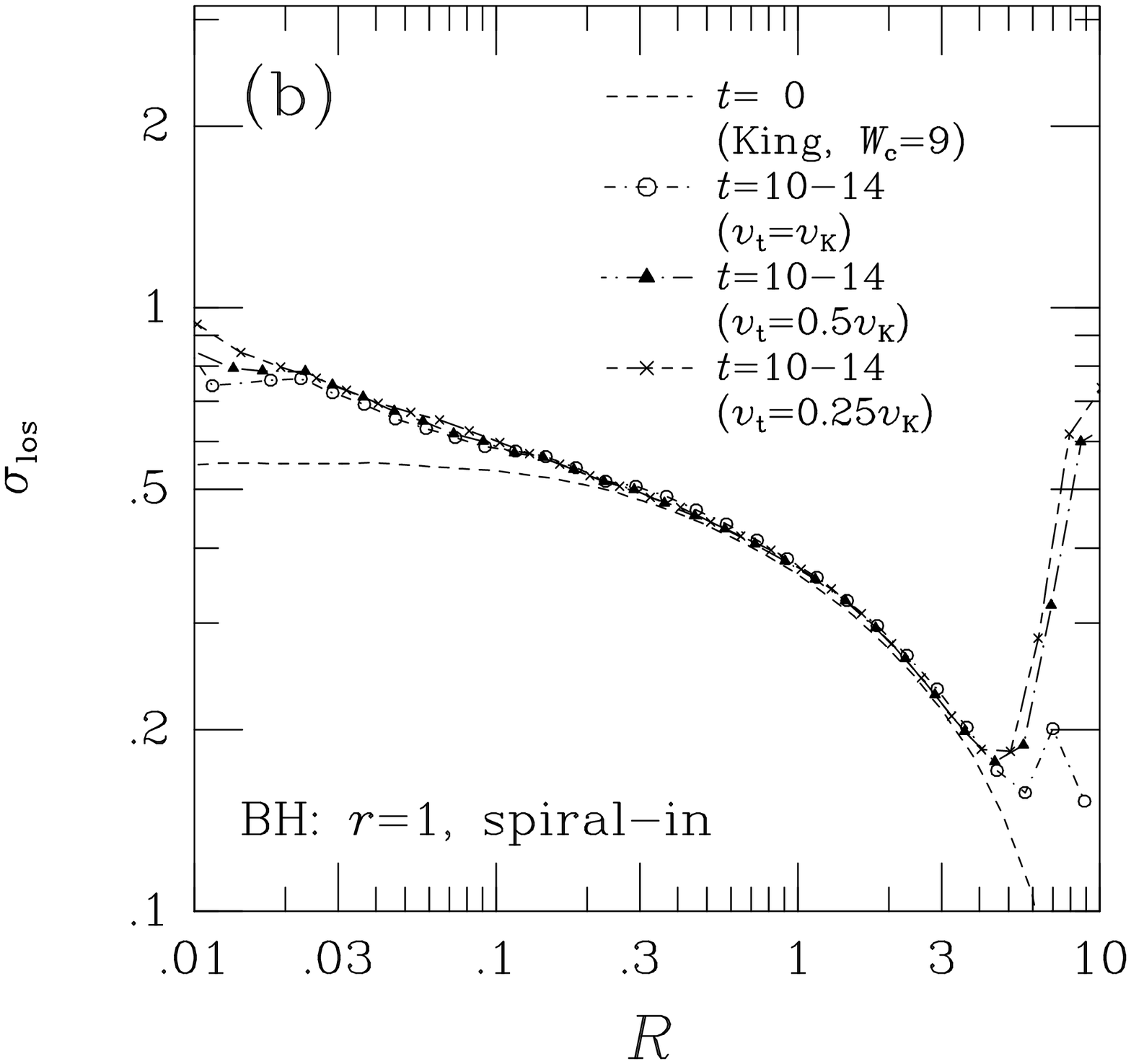}\\
\plotone{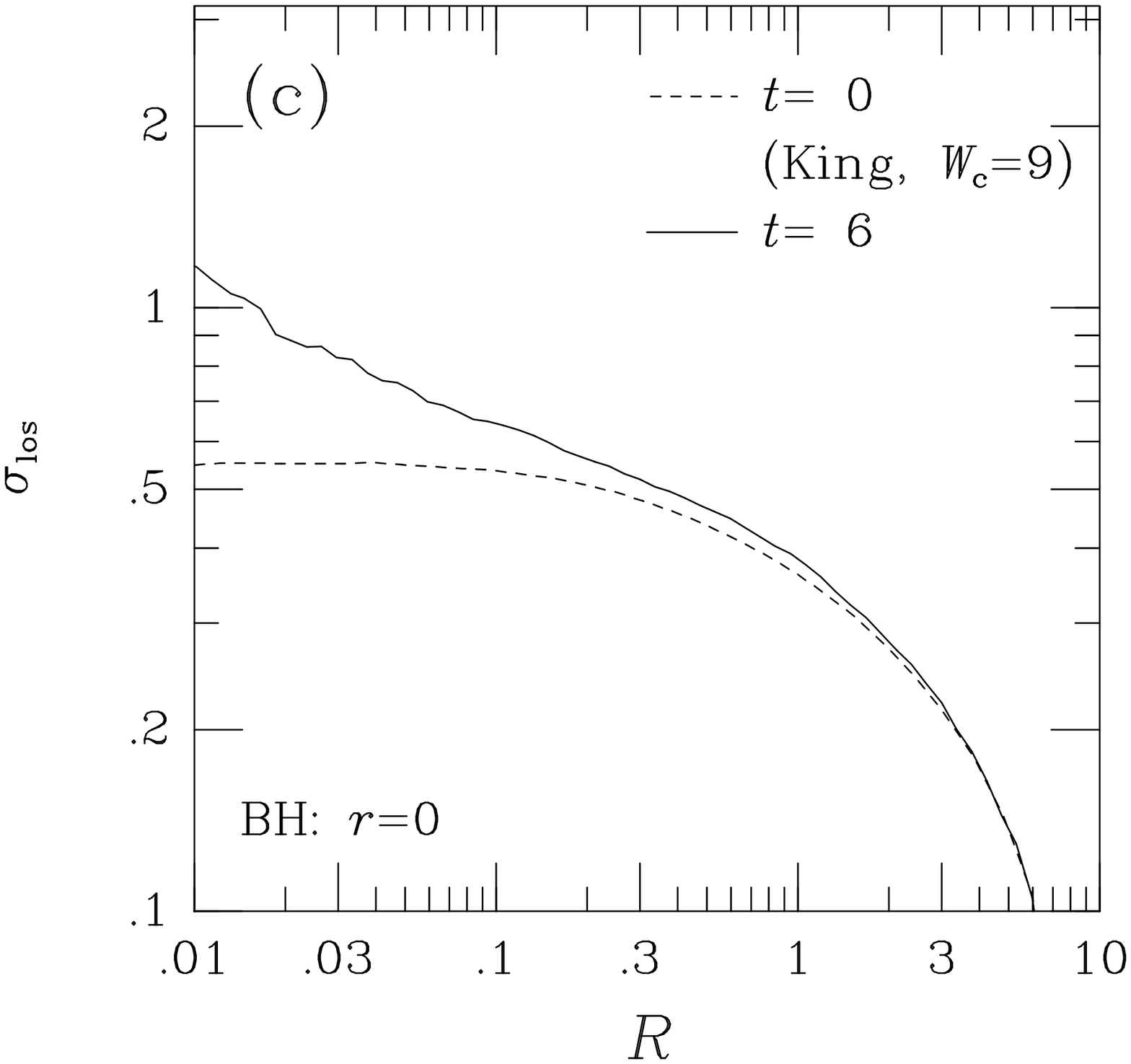}
\caption{Line-of-sight velocity dispersion profiles. (a) for %
off-center BH (run A); (b) for spiral-in BH (runs B, C and D); %
(c) for on-center BH (run E). \label{fig:7}}
\end{figure}

\begin{figure}[htbp]
\plotone{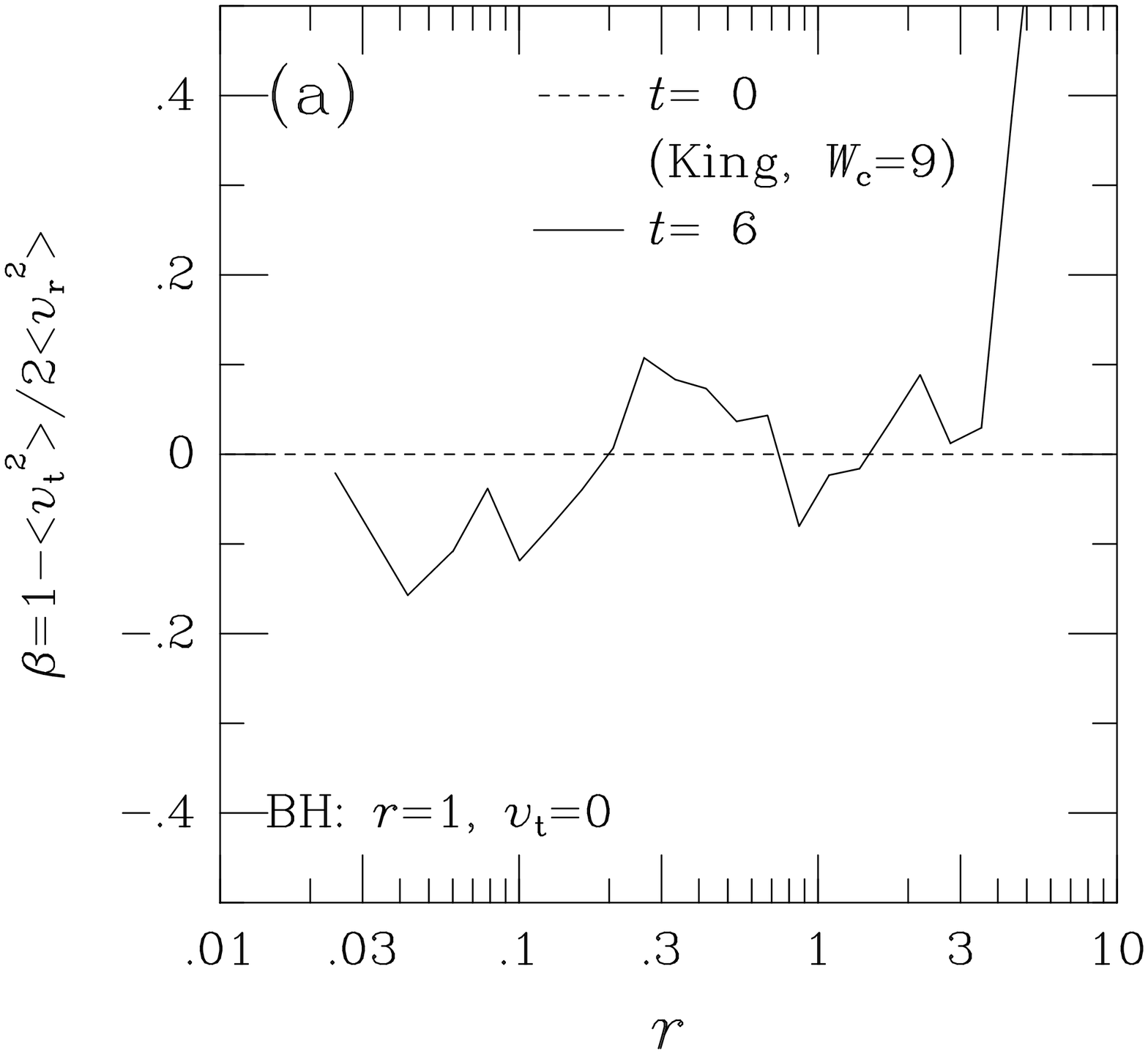}\\
\plotone{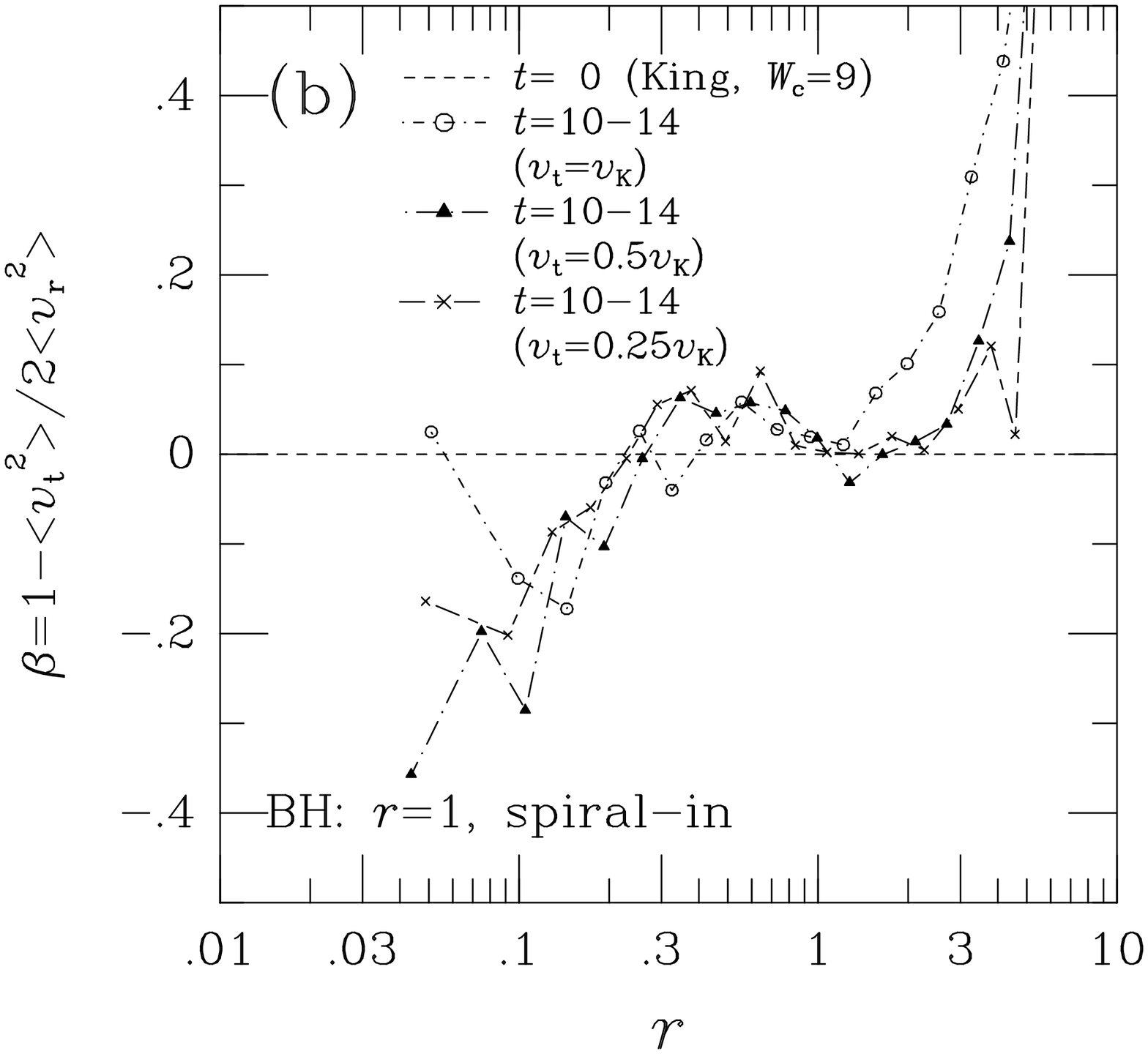}\\
\plotone{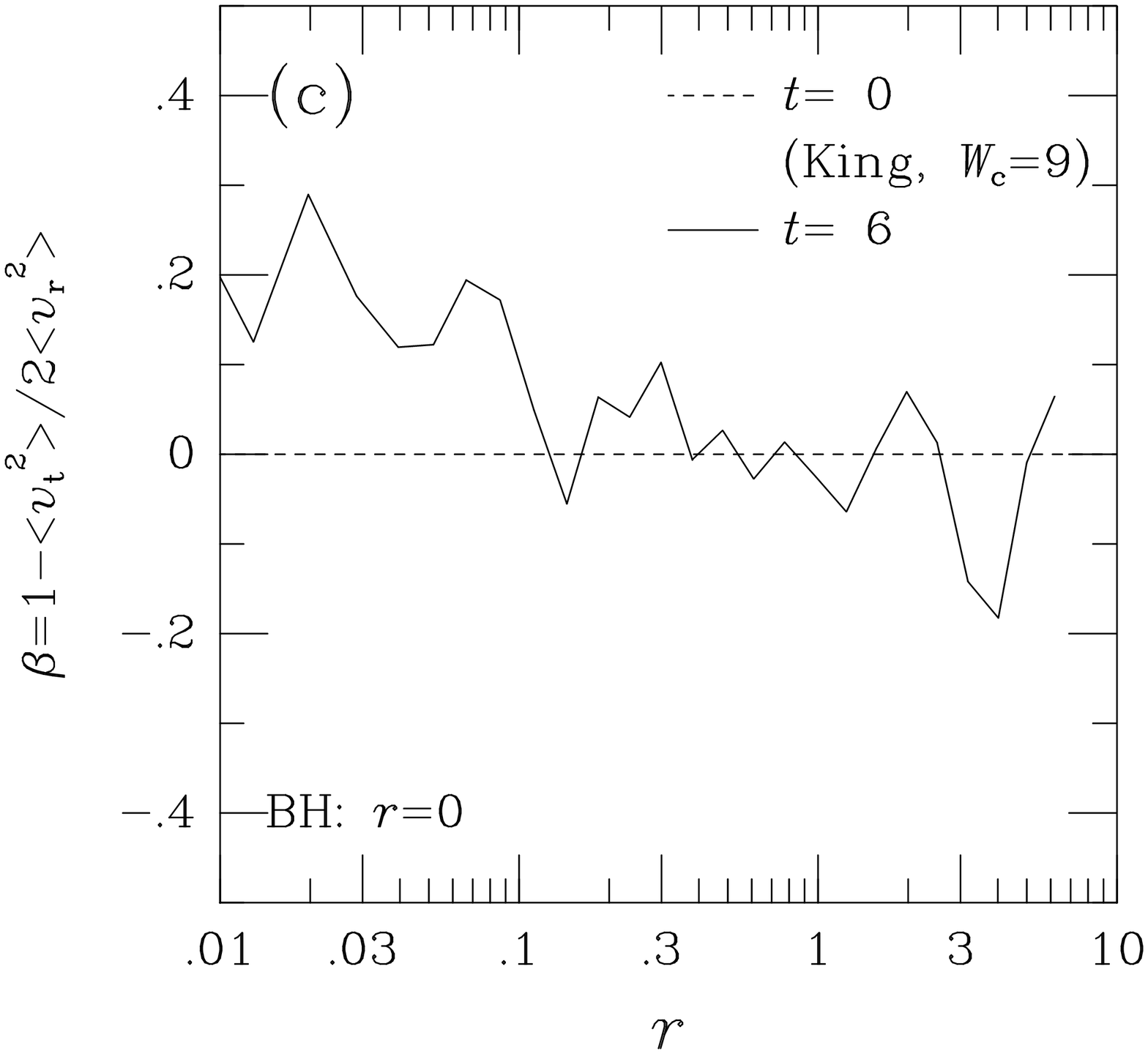}
\caption{Profiles of anisotropy parameter for the velocity %
dispersion. (a) for off-center BH (run A); (b) for spiral-in BH %
(runs B, C and D); (c) for on-center BH (run E). Plotted points %
are the averaged points of the data in logarithmic bins for $r$. %
We put 10 bins per decade. \label{fig:8}}
\end{figure}

\begin{figure}[htbp]
\plotone{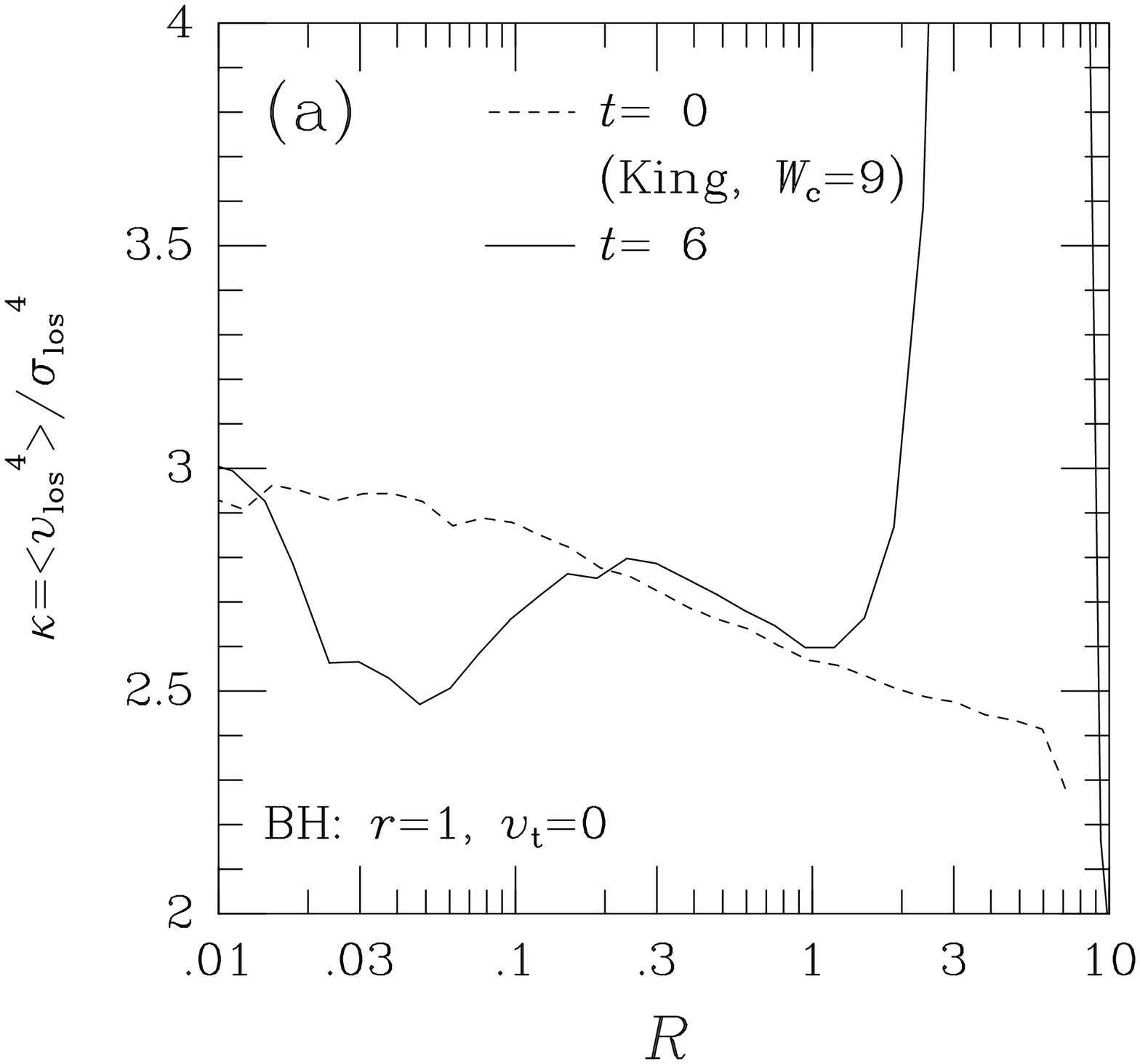}\\
\plotone{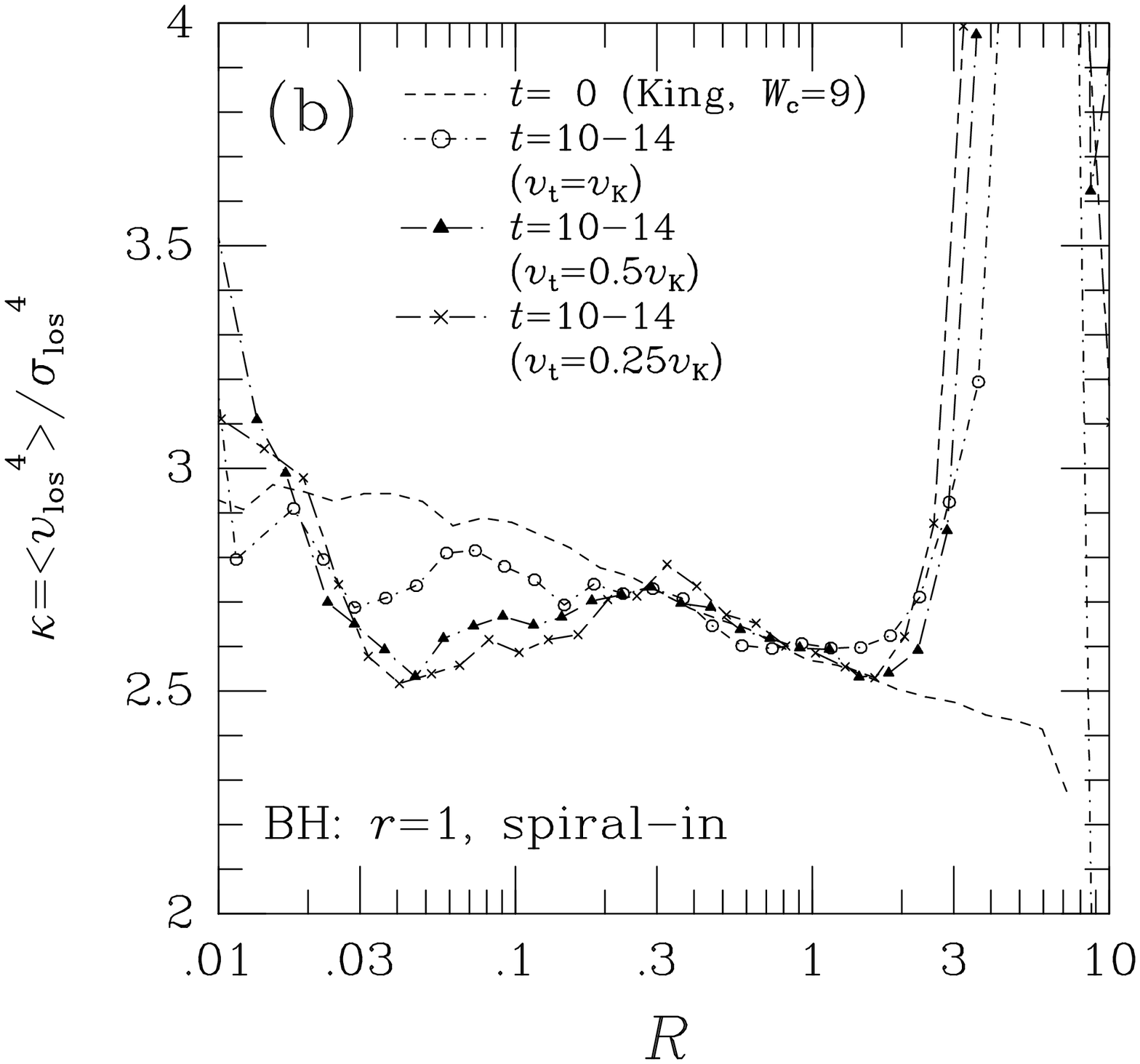}\\
\plotone{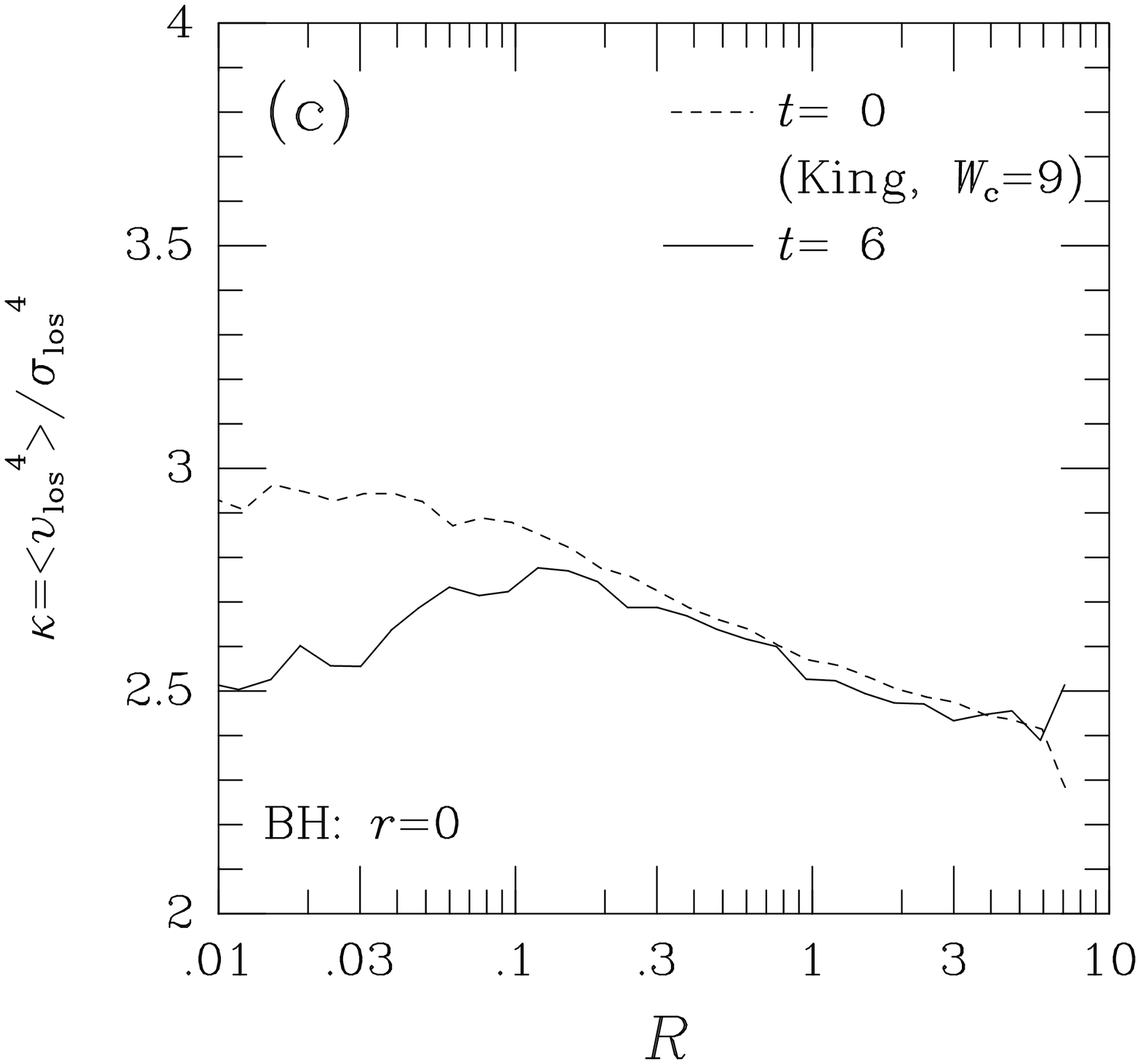}
\caption{Profiles of kurtosis of the line-of-sight velocity. %
(a) for off-center BH (run A); (b) for spiral-in BH %
(runs B, C and D); (c) for on-center BH (run E). Plotted points %
are the averaged points of the data in logarithmic bins for $r$. %
We put 10 bins per decade. \label{fig:9}}
\end{figure}

In Figures \ref{fig:6}, \ref{fig:7}, \ref{fig:8} and
\ref{fig:9}, we show the velocity profile, anisotropy and kurtosis.

The projected (line-of-sight) velocity dispersion is calculated using
the following formula (\cite{mer90}):
\begin{eqnarray}
\sigma_{\rm los}(R) &=& \left\{ \frac{2}{\Sigma (R)}\int_{R}^{\infty}
\frac{dr \; r\rho}{\sqrt{r^{2}-R^{2}}} \right. \nonumber \\%
 & & {} \left. \times \left[ \left( 1-\frac{R^2}{r^2}\right)
\langle v_{r}^{2} \rangle +\frac{1}{2}\frac{R^2}{r^2} \langle
v_{t}^{2} \rangle \right] \right\}^{1/2} \nonumber \\
 &=& \left\{ \frac{1}{2 \pi \Sigma (R)}\sum_{i(r_{i}>R)}
\frac{m_{i}}{r_{i}\sqrt{r_{i}^{2}-R^{2}}} \right. \nonumber \\%
 & & {} \left. \times \left[ \left
( 1-\frac{R^2}{r_{i}^2}\right) v_{r i}^{2} +
\frac{1}{2}\frac{R^2}{r_{i}^{2}} v_{t i}^{2} \right] \right\}^{1/2}, 
\end{eqnarray}
where $\langle v_{r}^{2} \rangle$ and $\langle v_{t}^{2} \rangle$ are, 
respectively, the radial and tangential mean square velocities.

In the central region ($r < 0.1$), the velocity dispersion is $\sigma
\sim r^{-1/2}$. This is simply because the gravitational potential is
dominated by the central black hole. In Figure \ref{fig:7}, we can
see that the projected one-dimensional velocity dispersion also shows
a marked increase at the center. The velocity dispersion profiles are
rather similar for off-center run and on-center run, since the central
potential is dominated by the black hole in both runs. However, if we
compare Figures \ref{fig:7}a and \ref{fig:7}c closely, it is
clear that the central increase of $\sigma_{\rm los}$ is less
pronounced in the case of the off-center run (Figure
\ref{fig:7}a). That is because the density cusp is weaker. The
results of the spiral-in runs (Figure \ref{fig:6}b and
\ref{fig:7}b) are almost the same as that of the off-center run,
like the result for density profile.

In Figure \ref{fig:8}, the velocity anisotropy parameter
\begin{equation}
\beta = 1-\langle v_{t}^{2} \rangle / 2 \langle v_{r}^{2} \rangle .
\end{equation}
is plotted against the distance from the center of the galaxy. From
this equation, $\beta = 0$, $-\infty$, $+1$ for an isotropic velocity
distribution, purely circular orbit, and purely radial orbit,
respectively.

Here, the difference between off-center and on-center runs is quite clear.

In run E, the change in the anisotropy is
monotonic, and inner region is more radial. This can be interpret as
the instantaneous response of stars to the injection of the central
black hole. If the central potential suddenly deepens, the orbit becomes more
radial. Note that this is quite different from the adiabatic response
of orbit studied by Young (1980) and Quinlan et al. (1995). In
the case of the adiabatic response, orbits become more circular.

In run A, however, the behavior of the anisotropy
is more complex. In the outermost region, orbits are predominantly
radial because the stars kicked out by the black hole populate the
outer halo. The effect of the black hole explains the general increase
in $\beta$ in the outer region.

However, in the innermost region, $\beta$ decreased. There are at
least two possible mechanisms for this decrease. First, the black hole 
should have ejected stars with radial orbit more efficiently than
those with circular orbit. Second, the increase in mass at the center of
the galaxy is not instantaneous as in the case of run E. Therefore,
the response of orbits would be somewhere in between increase in $\beta$
for run E and decrease in $\beta$ for the adiabatic response. If the
response is closer to adiabatic, it is natural that $\beta$ decreases.

In the spiral-in runs (runs B, C and D), $\beta$ shows the same
tendency as that in the off-center run: $\beta$ increased in outer
region and decreased in  inner region. The stars in central region of
the spiral-in run are somewhat more circular than those in the same
region of run A. Since the black hole deposited the angular momentum
to the field stars, this  result is quite natural.

Figure \ref{fig:9} shows the distribution of kurtosis, i.e. the
fourth moment of the line-of-sight velocity dispersion. We plot the
dimensionless kurtosis $\kappa=\langle v_{\rm los}^{4} \rangle /
\sigma_{\rm los}^{4}$ against the radius for runs A, B, C, D and
E. Here the kurtosis is calculated as follows (\cite{mer90}):
\begin{eqnarray}
\langle v_{\rm los}^{4} \rangle (R) &=& \frac{2}{\Sigma (R)}\int_{R}^{\infty}
\frac{dr \; r\rho}{\sqrt{r^{2}-R^{2}}} \nonumber \\
 & & \mbox{} \times \left[ \left( 1-\frac{R^2}{r^2}\right) ^{2} \langle v_{r}^{4} \rangle \right. \nonumber \\%
 & & {} \left. +3\frac{R^2}{r^4}(r^2-R^2) \langle v_{r}^{2} v_{t}^{2} \rangle +\frac{3}{8}\frac{R^4}{r^4} \langle v_{t}^{4} \rangle \right] \nonumber \\%
 &=& \frac{1}{2 \pi \Sigma (R)}\sum_{i(r_{i}>R)}
\frac{m_{i}}{r_{i}\sqrt{r_{i}^{2}-R^{2}}} \nonumber \\
 & & \mbox{} \times \left[ \left( 1-\frac{R^2}{r_{i}^2}\right) ^{2} v_{r i}^{4} \right. \nonumber \\%
 & & {} \left. + 3\frac{R^2}{r_{i}^4}(r_{i}^2-R^2) v_{r i}^{2} v_{t i}^{2} + \frac{3}{8}\frac{R^4}{r_{i}^4} v_{t i}^{4} \right]. 
\end{eqnarray}
In these figures, $\kappa=3$ corresponds to a Maxwellian velocity
distribution. Larger values of $\kappa$ ($\kappa > 3$) imply
high-velocity tails, and values of $\kappa < 3$ imply ``boxier''
distributions.

In all cases, the kurtosis is always less than 3 and decreases toward
the center, except in the very central region and in the outer
region. This result is consistent with the result for an adiabatic
growth BH model by Quinlan et al. (1995).

\placefigure{fig:6}

\placefigure{fig:7}

\placefigure{fig:8}

\placefigure{fig:9}

\subsection{Black Hole Mass \& Core Radius}
\begin{figure}[htbp]
\plotone{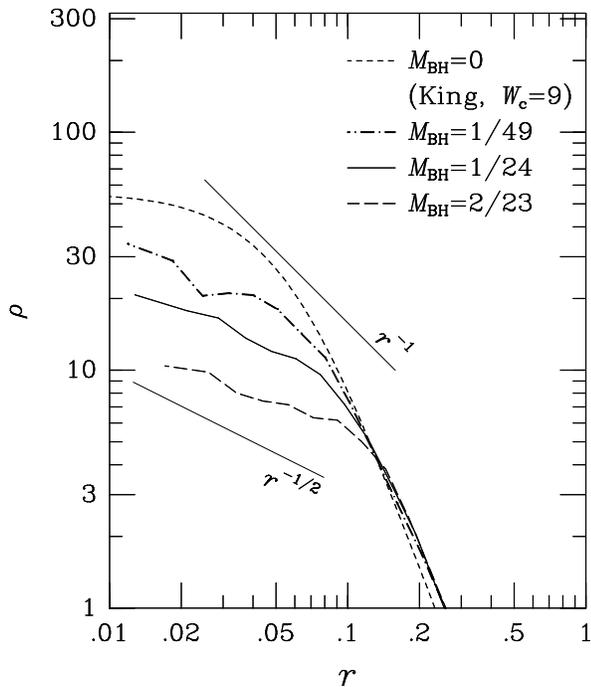}
\caption{Density profiles for BH mass $M_{\rm BH}=1/49, 1/24$ and %
$2/23$ (run F, A and G), respectively. \label{fig:10}}
\end{figure}

Figure \ref{fig:10} shows the time-averaged density profiles for
runs A, F and G ($M_{\rm BH} = 0.04, 0.02, 0.08 M_{\rm tot}$) for
$t=6-10$. We can see that more massive black holes make larger cusp
regions. We can also see that the slope of the cusp is almost the same
for all three runs.

\begin{figure}[htbp]
\plotone{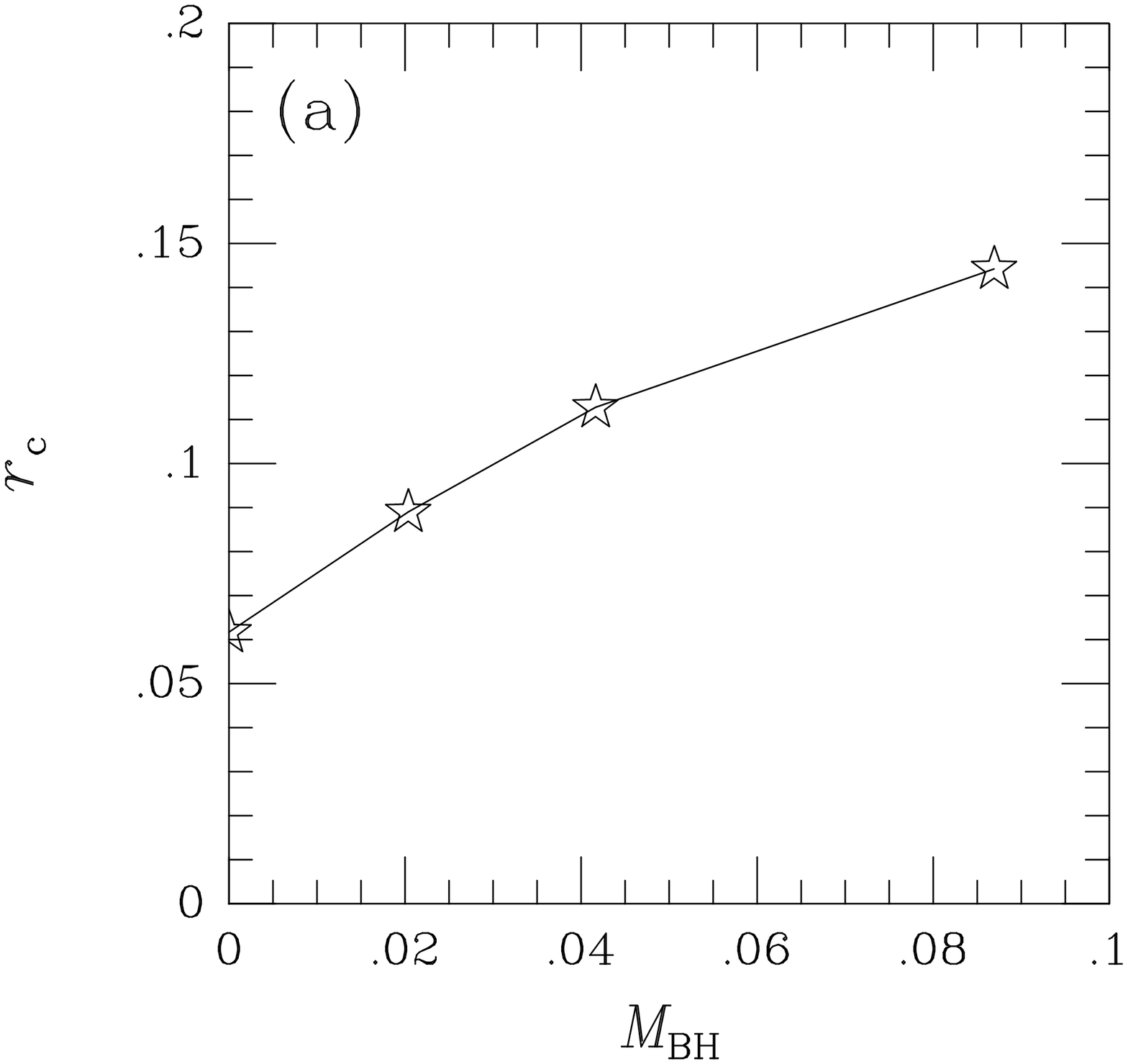}\\
\plotone{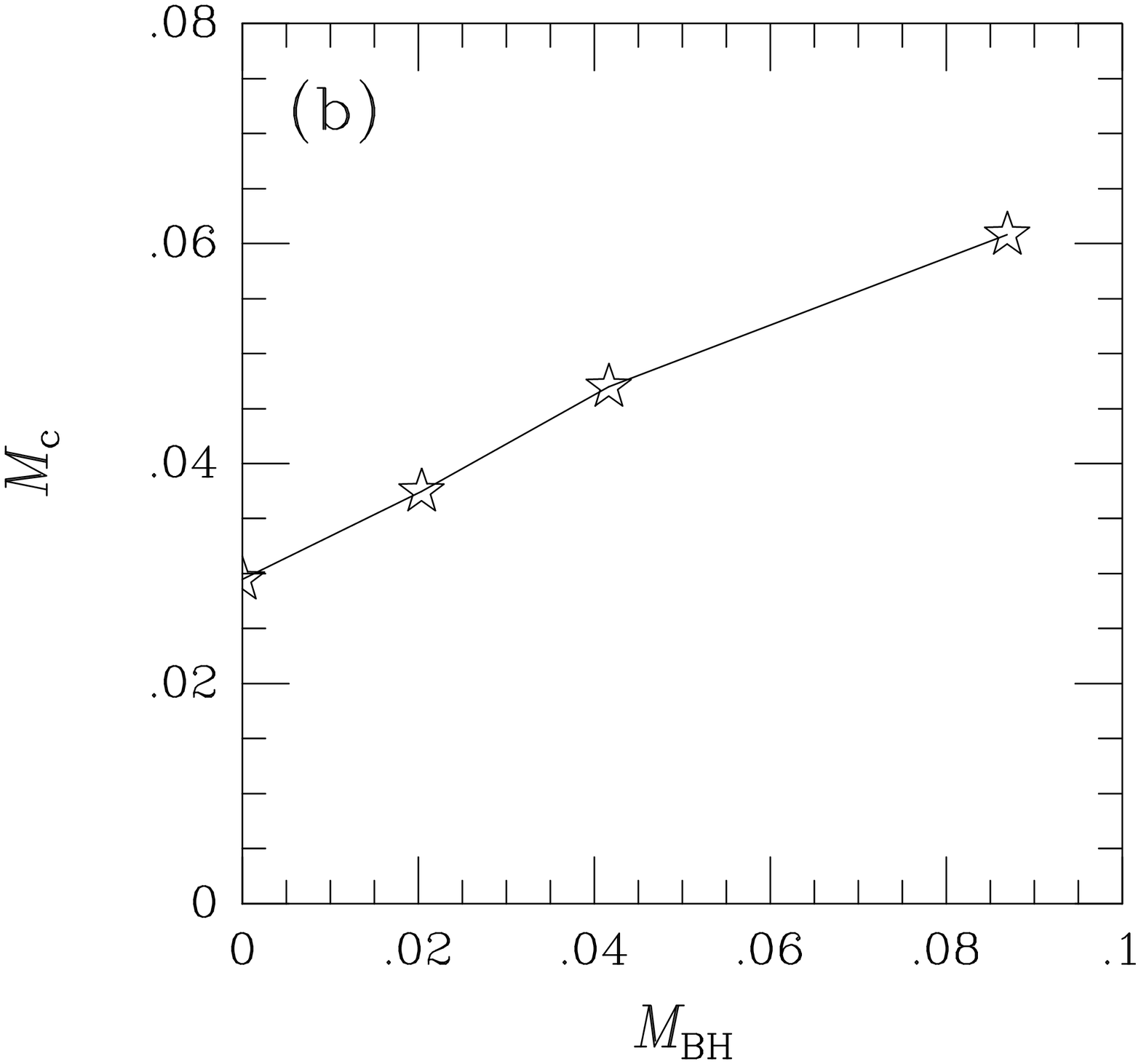}
\caption{(a) Relation between the BH mass $M_{\rm BH}$ and core radius %
$r_{\rm c}$ for the run F, A and G, respectively. (b) Relation between the %
BH mass $M_{\rm BH}$ and core mass $M_{\rm c}$ for the run F, A and G, %
respectively. \label{fig:11}}
\end{figure}

Figure \ref{fig:11}a shows the relation between the black
hole mass and the ``core radius''. We adopted the
definition of the core radius by McMillan, Hut, \& Makino (1990):
\begin{equation}
r_{\rm c}=\sqrt{\frac{\sum_{i}^{} \rho_{i}^{2}r_{i}^{2}}{\sum_{i}^{}
\rho_{i}^{2}}},
\end{equation}
where $r_{i}$ is the distance of star {\it i} from the central black
hole and $\rho_{i}$ is the local density at the position of star {\it
i}. This definition gives a good estimate of the size of the ``weak
cusp'' region (\cite{mak96}). Figure \ref{fig:11}b shows
the relation between the black hole mass and the mass of stars within
the core radius. 

Figures \ref{fig:11}a and b indicate that the core size shows almost
linear dependence on the black hole mass. Of course, the relation is
not strictly linear simply because the initial King model had a finite
core. Even for zero black hole mass the core size is still finite.

\placefigure{fig:10}

\placefigure{fig:11}

\subsection{Dependence on the Galaxy Model}
\begin{figure}[htbp]
\plotone{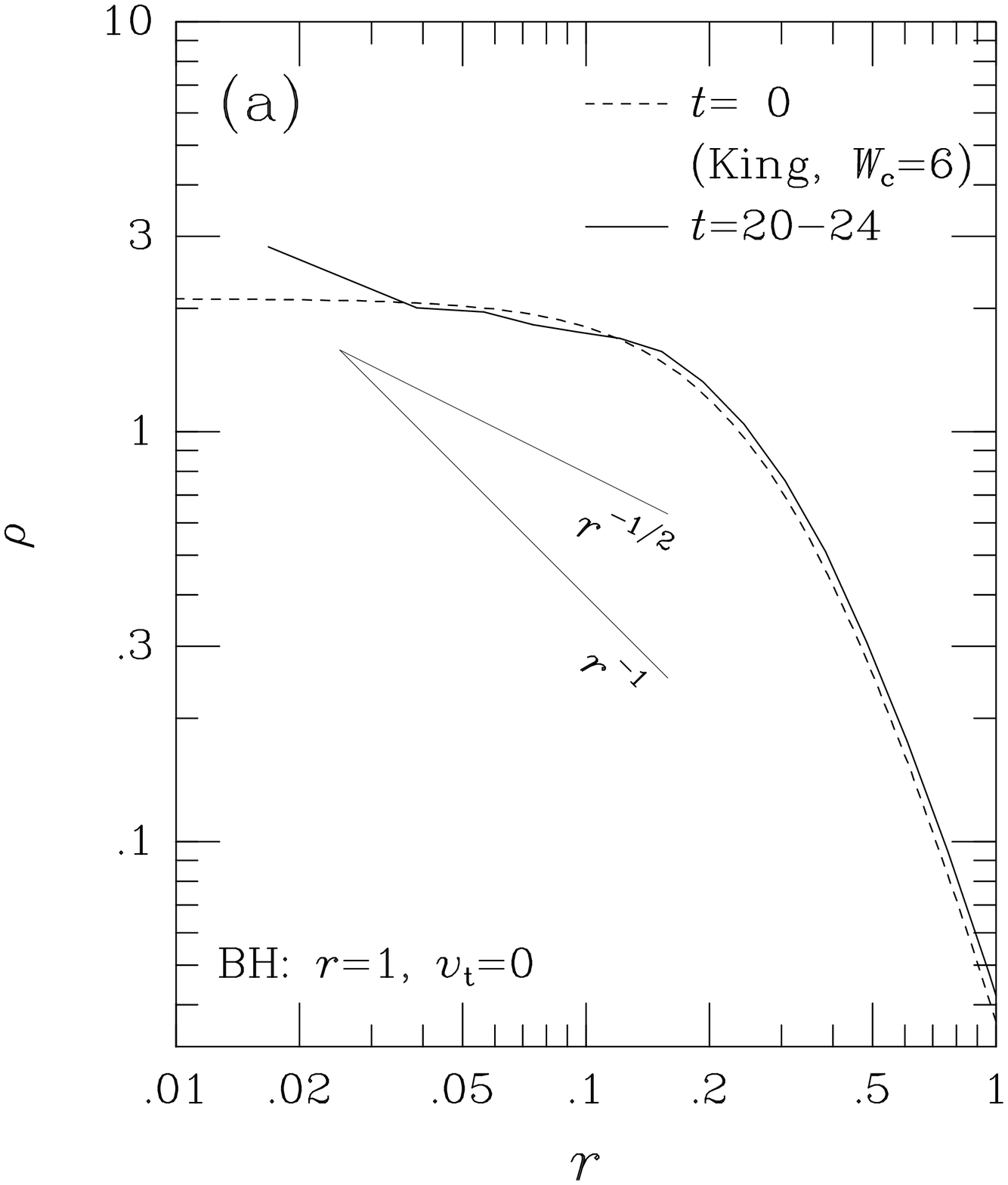}\\
\plotone{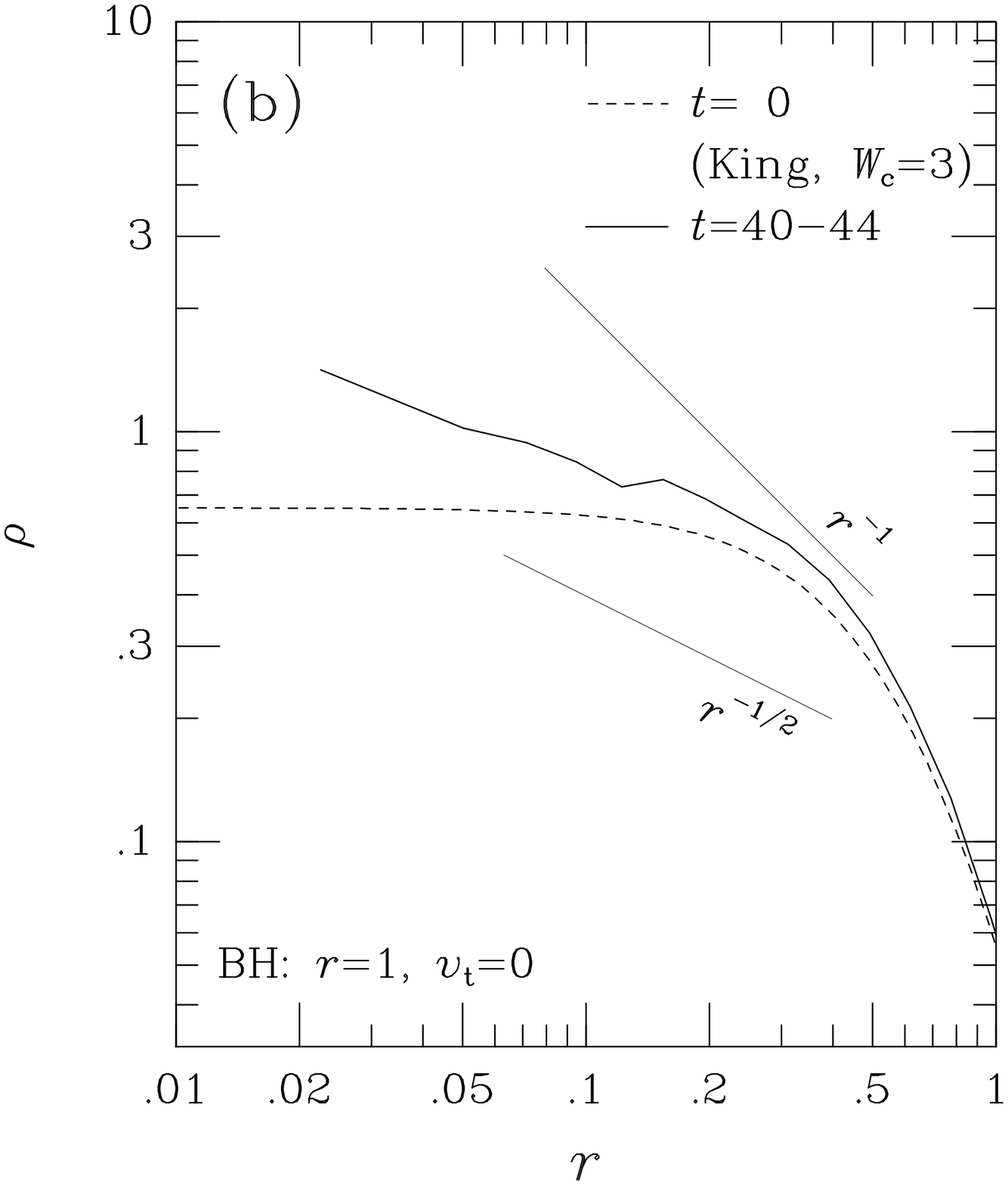}
\caption{(a) Density profile for King model with $W_{\rm c}=6$ (run %
H). (b) Same as (a) but for King model with $W_{\rm c}=3$ (run %
I). \label{fig:12}}
\end{figure}

Figure \ref{fig:12} shows the time-averaged density profile for
$t=20-24$ for run H and $t=40-44$ for run I. We used less concentrated
King model ($W_{\rm c}=6, 3$) as initial galaxy models for these
runs. It has the core about 10 times larger than that of the model we
used for all other runs. The result of run H (figure \ref{fig:12}a)
shows that the structure of the core remains almost unchanged. The
result of run I (figure \ref{fig:12}b) shows that the flat core is
replaced by a slight increase in the central density ($\rho \propto
r^{-0.3}$). However, in this run, the initial potential energy of the
black hole is not much larger than the bound energy of the stars in
the core, since the core size is as large as the initial distance of
the black hole from the center of the galaxy. Therefore, this
``cuspy'' profile is likely to be formed through the process similar
to that for run E, when we placed the black hole in the core.

\placefigure{fig:12}

\subsection{Effect of Two-Body Relaxation}
\begin{figure}[htbp]
\plotone{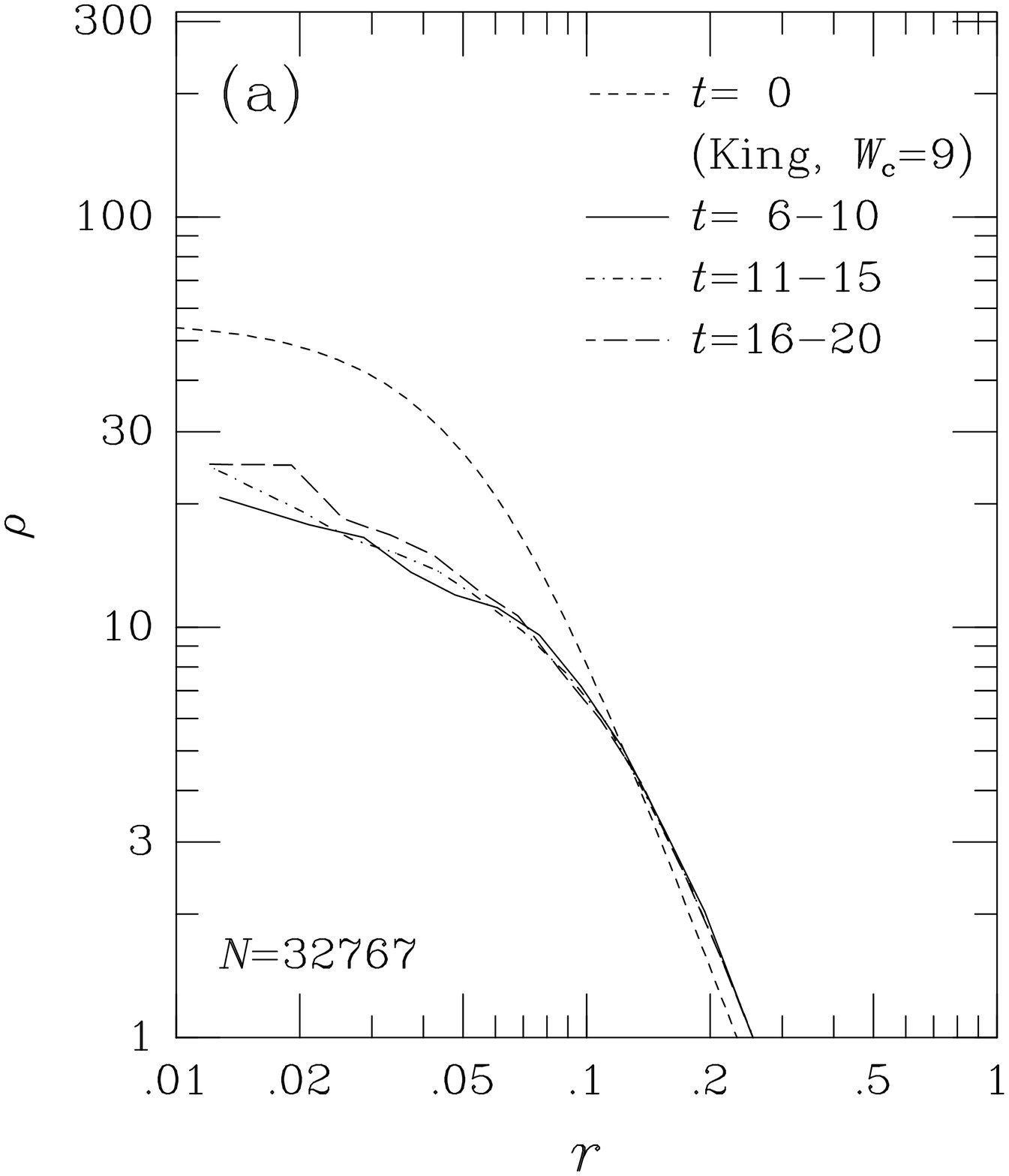}\\
\plotone{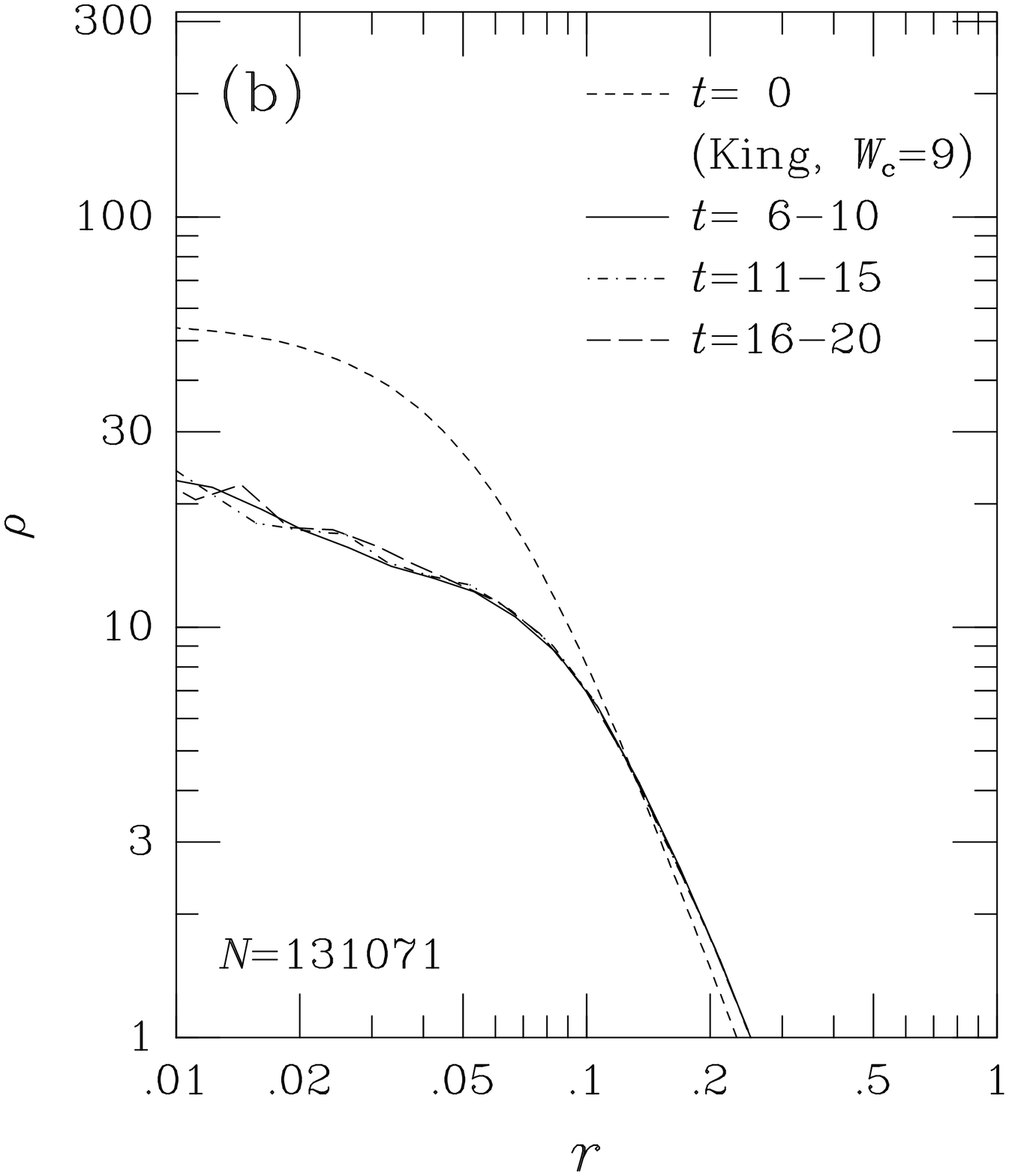}
\caption{(a) Time evolution of density profile for $N=32767$ (run %
A). (b) Same as (a) but for $N=131071$ (run J). \label{fig:13}}
\end{figure}

Figure \ref{fig:13} shows the time evolution of the density profile
for runs with different number of particles (runs A and J). All other
parameters are the same. In Figure \ref{fig:13}a, it is clear that the 
central cusp becomes steeper as the system evolves, while in Figure
\ref{fig:13}b (run J) the change is negligible. This difference is due 
to the difference in the thermal relaxation time. The model in run A
has much shorter relaxation time. Real galaxies have the relaxation
time several orders of magnitude larger than that of the model in run
J. Therefore, in real ellipticals the thermal evolution would be
negligible and the profile would not change.

\placefigure{fig:13}

\section{Discussion and Summary} \label{sec:summary}

We investigated the dynamical effect of a massive black hole which
sinks into the galactic center to the structure of the galaxy by means
of $N$-body simulations. We obtained three main results.

First, we found that the black hole destroys the central core of the
initial galaxy by heating up the field stars. After the black hole
settled in the center, a weak density cusp ($\rho \sim r^{-0.5}$)
remained around the central black hole. This result does not depend on
how much angular momentum the sinking black hole initially has. This
result nicely explains the recent {\it HST \/}observations of large
elliptical galaxies (\cite{lau95}; \cite{byu96};
\cite{geb96}; \cite{fab96}; \cite{kor96}). On the other hand, when a
massive black hole is placed initially at the galactic center, the
central density increases and a steep cusp ($\rho \propto r^{-1.5}$)
is formed.

Second, the velocity anisotropy profiles are quite different between
the above two cases. For the off-center black hole model, the
orbits of the stars in the outer region tend to be radial, while in
the inner region, the orbits tend to be circular. If the black hole
spirals into the center, the field stars get angular momentum from the 
black hole and the orbits become more circular in the innermost
region. For the on-center black hole model, however, the velocity is
more radial near the center and approximately isotropic in the outer region. 

Third, the size of the weak cusp region shows a clear correlation with
the black hole mass. The heavier black holes form larger cores.

From the present results, we can conclude that the weak density cusp
can be formed if a massive black hole dynamically sinks to the center
of a galaxy. This suggests that the prime cause of
the formation of the weak cusp in simulations by ME is the back
reaction to the sinking black holes. Moreover, the origin of the weak cusp
in large elliptical galaxies found with recent {\it HST\/} observations
can be explained by this galaxy merging and subsequent sinking of
black holes. From this point of view, the dichotomy of elliptical
galaxies (``core'' galaxies and ``power-law'' galaxies (\cite{lau95};
\cite{byu96}; \cite{geb96}; \cite{fab96}; \cite{kor96})) could be
understood as follows: small and therefore ``cuspy'' ellipticals and
normal spirals are most likely to be formed through dissipational
collapse. During the formation process, they are likely to contain
moderately massive central black holes ($\sim 10^{7} M_{\odot}$). When
these galaxes merge, black holes spiral into the center of the merger
and left the weak cusp. If the merger still contain gas, significant
fraction of that mass would accrete to the black hole
(\cite{tan96}). In this process, ``cuspy'' profile might be
restored. However, if gas is already consumed by some process, the
weak cusp is preserved. Thus, massive ellipticals are likely to have
weak cusps, while less massive ones are likely to have deep cusps,
even if they are mergers.

In addition, we investigated the possibilities of identifying the
mechanisms of the formation of the central cusps in elliptical
galaxies by comparing our results with the observations. We calculated
the line-of-sight velocity dispersion and the kurtosis for both
models, but the difference in the profiles of them are quite
small. There was a clear difference in the anisotropy. However, to
determine anisotropy from observation is not easy. Thus, it will be
very difficult to detect the difference of the two types of the cusps
in the spectroscopic data. The most clear difference is in the density
profile. If the improved velocity dispersion data for the observed
``weak-cusp'' galaxies demonstrates the presence of massive black
holes, our scenario would be the most plausible explanation.

\acknowledgments

We thank Yoko Funato for helpful comments on the simulation method and
the manuscript. We are grateful to Shunsuke Hozumi for valuable
conversations and for his comments on previous drafts of this
paper. T.N. thanks Toshiyuki Fukushige for much useful advice on the
calculation method and for stimulating discussions. We also thank
Daiichiro Sugimoto and all the people who developed the
special-purpose computer GRAPE-4.

\end{document}